\documentclass[11pt]{article}
\usepackage{amsmath,amssymb}
\usepackage[margin=1in]{geometry}
\usepackage{xcolor}
\usepackage[colorlinks=true,linkcolor=blue,citecolor=blue,urlcolor=blue,hypertexnames=false]{hyperref}
\hypersetup{pdftitle={Exact Planar Black Hole in AdS-Einstein-Scalar Gravity with IR-Emergent Nearly Conformal Fluid},pdfauthor={Sangheon Yun}}
\usepackage{cite}
\usepackage[protrusion=true,expansion=true]{microtype}

\newcommand{\bea}{\begin{eqnarray}}
\newcommand{\eea}{\end{eqnarray}}
\newcommand{\be}{\begin{equation}}
\newcommand{\ee}{\end{equation}}
\def\rmd{{\rm d}}

\title{\bf Exact Planar Black Hole in AdS-Einstein-Scalar Gravity\\ with IR-Emergent Nearly Conformal Fluid\\[9mm]}
\author{Sangheon Yun\\[5mm]
{\small\it IndigoWave, Center for Quantum Spacetime, Sogang University,}\\
{\small\it 35 Baekbeom-ro, Mapo-gu, Seoul 04107, Republic of Korea}\\[4mm]
{\small\tt sangheon.yun@gmail.com}\\[9mm]}
\date{}

\begin{document}
\maketitle

\begin{abstract}
We study an exact static planar hairy black hole of Einstein gravity minimally coupled to a neutral scalar in four dimensions, which arises in a consistent truncation of type IIA supergravity on AdS$_4\times\mathbb{CP}^3$ dual to the ABJM theory.
For nonzero hair the asymptotic region is hyperscaling violating rather than AdS, and the dual theory is a three-dimensional QFT deformed in the UV by an effectively marginally relevant operator with logarithmic running, which flows to the ABJM fixed point in the IR.
On the gravity side, treating the hair parameter as an independent variable leads to a thermodynamic central charge $C_t$ and a generalized first law, organized by a radial scaling isometry.
Holographic renormalization by the Hamilton--Jacobi method gives the one-point functions in closed form and the trace Ward identity.
With the source held fixed, black branes exist below a maximal temperature $T_{\max}\simeq0.19\,\Lambda$ in two branches.
On the stable branch the plasma approaches the conformal ABJM fluid with corrections of relative order $(T/\Lambda)^3$, which arise from a state-dependent rescaling of the boundary coordinates, while the other branch is governed by hyperscaling violation with $\theta=9$.
A scale-invariant multi-trace boundary condition, which fixes the hair parameter instead of the source, selects the same solutions as the thermal states of a second theory, whose energy density is the bulk mass density and whose equilibrium thermodynamics is exactly conformal.
Since the source mode is not normalizable, this second theory is established only in equilibrium, and the theory with fixed source is the dual of the solutions.
\end{abstract}

\vspace{1.5cm}


\vspace{1.5cm}
\tableofcontents
\vspace{1.5cm}

%%%%%%%%%%%%%%%%%%%%%%%%%%
\section{Introduction}
%%%%%%%%%%%%%%%%%%%%%%%%%%

The gauge/gravity duality \cite{Maldacena:1997re,Gubser:1998bc,Witten:1998qj} provides a powerful framework for studying strongly coupled field theories through their dual gravitational descriptions, and compactifying higher-dimensional supergravities to four dimensions gives access to the strongly coupled regime of the resulting low-energy effective theory.
The four-dimensional effective action relevant to ABJM theory \cite{Aharony:2008ug} was obtained in \cite{Gauntlett:2009zw} from eleven-dimensional supergravity and in \cite{Bak:2010yd} from type IIA supergravity.
Here we focus on a minimal truncation of that action -- Einstein gravity minimally coupled to a single neutral scalar $\phi_+$ -- viewed as a minimal holographic model of thermal ABJM dynamics, in which the dominant thermal mode surviving the neutral truncation condenses.

Holographic hydrodynamics \cite{Policastro:2001yc,Kovtun:2004de} describes the long-wavelength dynamics of strongly coupled plasmas in terms of black brane horizons.
When the boundary theory is conformal the dual fluid is conformal as well, with speed of sound $c_s^2=1/(D_{bdy}-1)$ ($D_{bdy}$ is the dimension of the boundary spacetime, here $D_{bdy}=3$), vanishing bulk viscosity and a traceless stress tensor \cite{Baier:2007dw}.
How non-conformal UV physics is reflected in the transport properties of the plasma is a central question \cite{Kanitscheider:2009mn}, studied for instance in holographic RG flows triggered by classically marginal operators \cite{Park:2021,Bourdier:2013axa}, in non-conformal deformations of black brane spectra \cite{Saha:2025qnm}, and in bottom-up models designed to mimic the QCD equation of state \cite{Gubser:2008ny}.
In most such models the background is known only numerically.
Exact solutions of Einstein-scalar gravity with exponential potentials \cite{Cadoni:2011sm,Cadoni:2012uf,Yun:2010,Gouteraux:2011ce,Bremer:1998zp,Charmousis:2010zz} are therefore valuable, since they allow the full holographic dictionary to be worked out analytically.

An exact static planar hairy black hole (equivalently, a hairy black brane) of Einstein-scalar gravity was obtained in \cite{Cadoni:2011sm}, and its thermodynamics was analyzed in \cite{Cadoni:2012uf}, where hyperscaling violation in the UV and the restoration of conformal invariance in the infrared were observed.
These works did not relate the model to ABJM theory, although it was noted in \cite{Cadoni:2012uf} that the model also arises from the compactification of black 2-branes with a smeared charge \cite{Gouteraux:2011ce}.
Meanwhile, the identification of the same model as a consistent truncation of type IIA supergravity on $\mathrm{AdS}_4\times\mathbb{CP}^3$ dual to ABJM theory was established separately in \cite{Bak:2010yd}.
The present paper combines these two lines of work and develops the gravitational and holographic description of the solution.

First, we analyze the black hole thermodynamics on the space of solutions.
Treating the horizon parameter $m$ and the hair parameter $\alpha$ as independent, rather than relating them by a constraint that leaves a single parameter as in \cite{Cadoni:2012uf}, we find that the entropy at fixed temperature depends on $\alpha$ only through $e^{3\alpha/2}$, in direct analogy with the role of $N^{3/2}$ in the thermodynamics of ABJM plasma \cite{Bak:2010yd}.
We identify this combination as a thermodynamic central charge $C_t$, in terms of which the mass density obeys a generalized first law $\rmd M=T\,\rmd S+\mu\,\rmd C_t$ together with Euler and Smarr relations, with the same structure as the holographic thermodynamics of CFTs with a variable central charge \cite{Visser:2021eqk,Cong:2021fnf,Ahmed:2023snm}.
This structure is organized by a radial scaling isometry of the solution space.
Whether the two-parameter family of solutions admits a consistent thermodynamic interpretation was left as an open question in \cite{Cadoni:2012uf}; we answer it in the affirmative, the second parameter being related on the boundary to the source of the dual operator (Section~\ref{sec:holothermo}) or to a multi-trace coupling (Section~\ref{sec:multitrace}).

Second, we carry out the holographic renormalization of the solution.
Since the UV asymptotics is that of a non-conformal brane rather than AdS, we use the Hamilton--Jacobi approach \cite{Boer:2000,Kanitscheider:2008kd}, in which the counterterm is the fake superpotential of the zero-temperature domain wall \cite{Yun:2010,Bremer:1998zp}.
Since the general results of holographic renormalization assume AdS asymptotics, we state explicitly which parts of the dictionary we assume, and we show that the source mode, unlike the state-dependent mode, is not normalizable.
The one-point functions of the stress tensor and of the scalar operator follow in closed form.
The energy density differs from the mass density by a contribution of the scalar condensate, $\varepsilon=M+\frac{2}{\sqrt7}\langle\mathcal{O}_\phi\rangle$, and the trace of the stress tensor obeys the Ward identity $\langle T^i{}_i\rangle=-\beta_\infty\langle\mathcal{O}_\phi\rangle$, where $\beta_\infty$ is the UV value of the holographic beta function \cite{Boer:2000}.

Third, we formulate the thermodynamics of the dual field theory.
Its data are the flat boundary metric and the source $\varphi_{(0)}$ of $\mathcal{O}_\phi$, which defines a dynamically generated scale $\Lambda$; the source depends on the solution only through the combination $(\sinh\alpha)/m$, while $\alpha$ alone determines $T/\Lambda$.
The natural variables are therefore the entropy density and the source, or equivalently $T$ and $\Lambda$, and the holographic first law reads $\rmd\varepsilon=T\,\rmd s-\langle\mathcal{O}_\phi\rangle\,\rmd\varphi_{(0)}$.
Together with the Euler relation $\varepsilon+p=Ts$, the scaling of the free energy with $\Lambda$ reproduces the trace Ward identity, which is in turn the boundary image of the bulk Smarr relation.
In a given boundary theory $C_t$ becomes a temperature-dependent effective count of degrees of freedom that reduces to the ABJM value at low temperature.

Fourth, we determine the phase structure and the hydrodynamics of the dual plasma.
Black branes exist only below a maximal temperature $T_{\max}\simeq0.19\,\Lambda$ and come in two branches.
On the branch connected to the zero-temperature domain wall the specific heat is positive and the black brane dominates the canonical ensemble; as $T\to0$ the plasma approaches the conformal ABJM fluid, with $c_s^2\to1/2$ and, through the Eling--Oz formula \cite{Eling:2011ms}, $\zeta/\eta\to0$, because the horizon value of the scalar becomes independent of the entropy in this limit.
This explains the emergent infrared conformality observed in \cite{Cadoni:2012uf}, and we compute the leading corrections, of relative order $(T/\Lambda)^3$, analytically.
On the other branch the horizon lies in the hyperscaling-violating region and the thermodynamics is governed by the exponent $\theta=9$ \cite{Charmousis:2010zz,Dong:2012se}, with $s\propto T^{-7}$, negative specific heat and $c_s^2\to-1/7$, in agreement with the universal relations of non-conformal branes \cite{Kanitscheider:2009mn}.
Throughout, the shear viscosity takes the universal value $\eta/s=1/4\pi$ \cite{Kovtun:2004de}.

Finally, we show that the same solutions admit a second holographic interpretation.
Under the scaling isometry the source shifts additively, so that a Dirichlet condition breaks scale invariance, whereas a multi-trace boundary condition \cite{Witten:2001ua,Henneaux:2006hk,Papadimitriou:2007sj} relating the condensate to the cube of the scale, $\hat{\mathcal O}=\rmd(\hat k\Lambda^3)/\rmd\varphi_{(0)}$, preserves it and fixes $\alpha$.
With this boundary condition the energy density is the bulk mass density $M$, and the equilibrium thermodynamics is that of a conformal fluid at all temperatures, with $c_s^2=1/2$ and $\zeta=0$; the generalized first law of the bulk then describes a varying multi-trace coupling $\hat k$, of which $C_t$ is a function.
Because the source mode is not normalizable and the energy is unbounded below, this theory is established only in equilibrium.
We therefore regard the theory with fixed source as the dual of the solutions, and the fixed-$\hat k$ theory as an alternative whose equilibrium thermodynamics accounts for the conformal relations obeyed by the bulk mass density at fixed $\alpha$.

The remainder of this paper is organized as follows.
In Section~\ref{sec:solution} we briefly review planar black holes and present the exact solution.
Section~\ref{sec:thermo} is devoted to the black hole thermodynamics on the two-parameter space of solutions.
In Section~\ref{sec:holoren} we carry out the holographic renormalization, derive the trace Ward identity and state the working assumptions on the holographic dictionary.
Section~\ref{sec:holothermo} formulates the holographic thermodynamics of the dual field theory and determines its phase structure.
Section~\ref{sec:boundary} discusses the holographic RG flow and the hydrodynamics of the dual plasma, including a comparison of the UV regime with the generalized conformal structure of non-conformal branes (Section~\ref{sec:KScomparison}), and Section~\ref{sec:multitrace} the scale-invariant boundary condition and the conformal fluid.
We conclude with a discussion and future directions in Section~\ref{sec:discussion}.
Details of the near-boundary analysis are collected in Appendix~\ref{app:holoren}, and homogeneous perturbations are analyzed in Appendix~\ref{app:dynamics}.

%%%%%%%%%%%%%%%%%%%%%%%%%%%%%%%%%%%%%%%%%%%%%%%%%%%%%%%%%%%%%%%%%%
\section{Hairy planar black hole}
\label{sec:solution}
%%%%%%%%%%%%%%%%%%%%%%%%%%%%%%%%%%%%%%%%%%%%%%%%%%%%%%%%%%%%%%%%%%

Throughout this paper a planar black hole, i.e.\ one whose horizon has flat transverse sections, is the same object as a black brane, and we use the two names interchangeably.
We consider Einstein gravity minimally coupled to a single neutral scalar field in four dimensions, together with the static ansatz of planar symmetry,
\bea
I&=&\frac{1}{16\pi G_4}\int\rmd^{4}x\sqrt{-g}\;
    \bigg[\mathcal{R}-\frac{1}{2}(\nabla\phi)^2 - V(\phi) \bigg]\,,\\
ds^2&=&-e^{2C(u)}\rmd t^2+e^{2B(u)}\rmd u^2
        +e^{2H(u)}(\rmd x^2+\rmd y^2)\,,
\eea
in which the scalar field depends only on the radial coordinate, $\phi=\phi(u)$.
Inserting the ansatz into the action gives the reduced action
\bea
I&=&\frac{1}{8\pi G_4}\!\int\!\rmd t\,\rmd u\,\rmd x\,\rmd y
   ~\mathrm{L}\,, \\
\mathrm{L}&=&e^{C - B + 2 H}\bigg[ 2C^\prime H^\prime + {H^\prime}^2
             - \frac{1}{4}{\phi^\prime}^2
             - \frac{1}{2}V(\phi)e^{2B} \bigg] \nonumber\\
             &-&\bigg[e^{C - B + 2 H}(C^\prime + 2 H^\prime)\bigg]'
\eea
where $'$ denotes $\frac{\rmd}{\rmd u}$.
One can then derive the equations of motion and use the definition of $u$ to set $B=C+2H$, yielding
\bea
&& C'' = H'' = 2 C^\prime H^\prime +{H^\prime}^2
   - \frac{1}{4}{\phi^\prime}^2
   = - \frac{1}{2}V(\phi)e^{2(C+2H)}\,, \nonumber \\
&& \phi'' = \frac{\rmd V(\phi)}{\rmd\phi}e^{2(C+2H)}\,.
   \label{phi}
\eea
For certain specific potentials the equations of motion admit exact solutions \cite{Cadoni:2011sm,Cadoni:2012uf,Yun:2010,Gouteraux:2011ce,Bremer:1998zp}.

To derive our exactly solvable hairy black brane configuration, we consider the minimal truncation of type IIA supergravity to Einstein gravity minimally coupled to a single neutral scalar $\phi_+$ \cite{Yun:2010},
\bea
\label{action}
I &=& \frac{1}{16\pi G_4}\int\rmd^{4}x\sqrt{-g}\;
      \bigg[\mathcal{R}-\frac{1}{2}(\nabla\phi_+)^2 - V(\phi_+)\bigg]\,,
      \nonumber \\
V(\phi_+) &=& - \frac{21}{2 l^2}e^{-3\phi_+/\sqrt{7}}
              + \frac{9}{2 l^2} e^{-\sqrt{7}\phi_+}\,,
\eea
where $l$ is the AdS$_4$ radius and we use natural units $c=\hbar=k_B=1$.
The potential satisfies $V(0) = -6/l^2$ (reproducing $2\Lambda = -6/l^2$) and $V''(0) = 18/l^2$, giving scalar mass squared $m_\phi^2 = 18/l^2$ at the $\phi_+=0$ vacuum (the pure AdS-Schwarzschild limit).
Equivalently, the operator $\mathcal{O}_\phi$ dual to $\phi_+$ has a conformal dimension $\Delta=6$ at this fixed point (from $\Delta(\Delta-3)=m_\phi^2 l^2=18$), hence it is \emph{irrelevant} at the $\phi_+=0$ (ABJM) point.
The solutions constructed below interpolate between this fixed point and a different asymptotic regime.
Towards the boundary the scalar runs off to $\phi_+\to\infty$, where the effective mass squared $m^2_{\rm eff}=V''(\phi_+)\to 0^-$ and the effective conformal dimension approaches the marginal value $\Delta_{\rm eff}\to 3^-$ (Section~\ref{sec:rg}); the ``marginally relevant'' characterization used throughout refers to this effective dimension in the large-$\phi_+$ region.
In the interior, the zero-temperature limit of the solutions flows to $\phi_+=0$, so that the ABJM fixed point is the IR end point of the flow and $\mathcal{O}_\phi$ is an irrelevant deformation there.
Recombining the equations of motion (\ref{phi}) gives \cite{Cadoni:2011sm}
\bea
&& C'' + 2 H'' - \frac{\sqrt{7}}{2}\phi_+''
   = \frac{9}{l^2} e^{2(C+2H-\sqrt{7}\phi_+/2)}\,, \nonumber \\
&& C'' + 2 H'' - \frac{3}{2\sqrt{7}}\phi_+''
   = \frac{9}{l^2} e^{2(C+2H-3\phi_+/2\sqrt{7})}\,, \nonumber \\
&& C'' = H'' \,, \nonumber \\
&& 2 C^\prime H^\prime +{H^\prime}^2 - \frac{1}{4}{\phi_+'}^2
   = \bigg(\frac{21}{4 l^2}e^{-3\phi_+/\sqrt{7}}
     - \frac{9}{4 l^2} e^{-\sqrt{7}\phi_+}\bigg) e^{2(C+2H)}\,,
\eea
which can be solved exactly in closed form.
After redefining $u$ and $t$, the explicit solution is
\bea
&& e^{C(u)} = \frac{l e^{2\delta u/3}}{3^{1/3}}
   \bigg( \frac{\gamma^{7} \sinh^{3}(\beta u + \alpha)}
               {\beta^{3} \sinh^{7}\gamma u}\bigg)^{\!1/12}\,, \nonumber \\
&& e^{H(u)} = \frac{e^{-\delta u/3}}{3^{1/3}}
   \bigg( \frac{\gamma^{7} \sinh^{3}(\beta u + \alpha)}
               {\beta^{3} \sinh^{7}\gamma u}\bigg)^{\!1/12}\,, \nonumber \\
&& e^{\phi_+(u)} = \bigg( \frac{\gamma \sinh(\beta u + \alpha)}
                               {\beta \sinh \gamma u} \bigg)^{\!\sqrt{7}/2}\,,
\eea
where $\alpha, \beta, \gamma$ and $\delta$ are constants satisfying
$\delta^2=\frac{7}{4}\gamma^2-\frac{3}{4}\beta^2$, with $\alpha\geq 0$.
In \cite{Cadoni:2011sm,Cadoni:2012uf} this class of solutions was presented in a more general form, for a family of potentials labelled by a parameter $\beta$; the potential in (\ref{action}) is the member $\beta^2=3/7$ of that family, with $\phi_+=-2\phi$ in the conventions of \cite{Cadoni:2012uf}.
Setting $\beta=\gamma=-\delta\equiv m>0$ and defining $r^3=ml^2(\coth mu +1)$, we obtain the exact static planar hairy black hole with scalar backreaction,
\bea
ds^2&=&-\bigg(\frac{r^2}{l^2}-\frac{2m}{r}\bigg)
        \bigg( \frac{\sinh \alpha}{m l^2}r^3 + e^{-\alpha} \bigg)^{1/2}
        \rmd t^2
      + \frac{\big( \frac{\sinh \alpha}{m l^2}r^3
               + e^{-\alpha} \big)^{3/2}}{r^2/l^2-2m/r}\,\rmd r^2
        \nonumber \\
&&+ r^2\bigg( \frac{\sinh \alpha}{m l^2}r^3 + e^{-\alpha} \bigg)^{1/2}
    (\rmd x^2+\rmd y^2) \nonumber \\
&\equiv& -{a(r)}^2\rmd t^2 + {b(r)}^2\rmd r^2
         + {h(r)}^2 (\rmd x^2+\rmd y^2)\,, \nonumber \\
e^{\phi_+} &=& \bigg( \frac{\sinh \alpha}{m l^2}r^3
               + e^{-\alpha} \bigg)^{\!\sqrt{7}/2}\,, \label{hairy}
\eea
where $t$, $x$, $y$ have been rescaled to absorb irrelevant factors, and $r_h \equiv (2ml^2)^{1/3}$ is the event horizon.
The solution is regular for $m>0$ and $\alpha\geq0$; we have verified directly that it solves the field equations following from (\ref{action}).
It is convenient to introduce the length $r_s$ defined by
\be
\label{rs}
r_s^3\equiv\frac{ml^2}{\sinh\alpha}\,,\qquad
\frac{\sinh\alpha}{ml^2}\,r^3+e^{-\alpha}=\Big(\frac{r}{r_s}\Big)^{3}+e^{-\alpha}\,,
\ee
which separates the region $r\ll r_s$, where $\phi_+$ is small and the geometry is close to AdS$_4$, from the scalar-dominated region $r\gg r_s$.
As shown in Section~\ref{sec:holoren}, $r_s$ is fixed by the source of $\mathcal{O}_\phi$ and therefore sets the intrinsic scale of the dual theory.
Since the solution is translationally invariant in the $(x,y)$ directions, all extensive thermodynamic quantities are defined per unit transverse area throughout this paper.
The Kretschmann scalar is finite at the horizon and at spatial infinity, confirming regularity outside the event horizon.
The only curvature singularity occurs at $r = 0$, hidden behind the horizon.

The near-horizon geometry for $m>0$ is the Rindler wedge times $\mathbb{R}^2$,
\be
\label{rindlerwedge}
ds^2= -\frac{3 r_h e^{\alpha/2}}{l^2}\,\rho\,\rmd t^2
      + \frac{l^2 e^{3\alpha/2}}{3 r_h \rho}\,\rmd\rho^2
      + r_h^2 e^{\alpha/2}(\rmd x^2+\rmd y^2)\,,
\ee
where $\rho = r-r_h$.
This confirms that our black brane is non-extremal.
Note that $\phi_+(r_h) = \frac{\sqrt{7}\alpha}{2}$, independent of $m$, since $(r_h/r_s)^3=2\sinh\alpha$.

The solution has several instructive limits.
For vanishing scalar field it reduces to pure AdS-Schwarzschild, and in the limit $m\to0$ with $\alpha/m\to0$ the metric reduces to AdS$_4$ in Poincar\'{e} coordinates.
When both $m$ and $\alpha$ tend to zero with $\sinh\alpha/m$ fixed, the metric yields the domain-wall solution of \cite{Yun:2010,Bremer:1998zp}: our black brane is therefore a thermal deformation of that zero-temperature soliton.
In terms of (\ref{rs}) the domain wall has $(r/r_s)^3+1$ in place of $(r/r_s)^3+e^{-\alpha}$: it approaches AdS$_4$ with $\phi_+=0$ as $r\to0$ and the hyperscaling-violating regime as $r\to\infty$, so it describes the zero-temperature RG flow from the deformed UV to the ABJM fixed point.
For $\alpha>0$ the region $r\gg r_s$ is not asymptotically AdS.
There $(r/r_s)^3$ dominates in (\ref{hairy}), and the metric becomes
\be
\label{UVmetric}
ds^2\simeq\Big(\frac{r}{r_s}\Big)^{3/2}r^2\Big(-\frac{\rmd t^2}{l^2}+\rmd x^2+\rmd y^2+\rmd\zeta^2\Big)\,,\qquad
\zeta\equiv2l\Big(\frac{r}{r_s^3}\Big)^{1/2}\,,
\ee
up to relative corrections of order $(r_s/r)^3$.
The geometry is conformally flat with a conformal factor growing as $\zeta^7$, which is the hyperscaling-violating form with $\theta=9$ and $z=1$ \cite{Charmousis:2010zz,Dong:2012se} discussed in Section~\ref{sec:rg}.
Its curvature vanishes instead of approaching the AdS$_4$ value $-12/l^2$: the trace of the field equations, $\mathcal R=\frac12(\nabla\phi_+)^2+2V$, gives $\mathcal R\simeq-\frac{105}{8l^2}(r_s/r)^{9/2}$.
The light-crossing time $\int b/a\,\rmd r$ grows as $r^{1/2}$, so that the asymptotic region is not a timelike conformal boundary.
These properties determine the status of the holographic dictionary used below (Section~\ref{sec:dictionary}).

Two further properties of the solution are used below.
The c-function \cite{Astefanesei:2007vh,Goldstein:2005hq}, $c(r) = \left[\frac{a(r)b(r)}{h'(r)}\right]^2$, increases monotonically with $r$, as shown in eq.~(\ref{cbeta}) of Section~\ref{sec:rg}, and is constant for $\alpha=0$.
Moreover, the scalar potential admits two fake superpotentials \cite{Elvang:2007ba,Yun:2010},
\bea
\label{fakeW}
&&V(\phi) = \frac{1}{2}\bigg[\bigg(\frac{\rmd W_\pm}{\rmd\phi}\bigg)^{\!2}
           - \frac{3}{4} W_\pm^2 \bigg]\,,\nonumber\\
&&W_\pm(\phi) = \frac{1}{l}\bigg(7e^{-3\phi/2\sqrt{7}}
          \pm 3e^{-\sqrt{7}\phi/2}\bigg)\,.
\eea
The lower sign, $W_-$, has a critical point at $\phi=0$ and generates the domain wall flowing to AdS$_4$; it defines the holographic renormalization scheme of Section~\ref{sec:holoren}.

%%%%%%%%%%%%%%%%%%%%%%%%%%%%%%%%%%%%%%%%%%%%%%%%%%%%%%%%%%%%%%%%%%
\section{Black hole thermodynamics}
\label{sec:thermo}
%%%%%%%%%%%%%%%%%%%%%%%%%%%%%%%%%%%%%%%%%%%%%%%%%%%%%%%%%%%%%%%%%%

The thermodynamics of solution (\ref{hairy}) was first analyzed in \cite{Cadoni:2012uf}, where a constraint relating the two parameters of the solution was imposed and the thermodynamic quantities were expressed in terms of a single parameter, $\nu_2=e^{-\alpha}$ in our notation; as shown in Section~\ref{sec:holothermo}, this constraint amounts to holding the source of the dual operator fixed.
In this section we keep both parameters $(m,\alpha)$ and describe the thermodynamics of the black brane on the two-parameter space of solutions.
Its translation into the thermodynamics of the dual field theory requires the holographic dictionary and is the subject of Sections~\ref{sec:holoren} and~\ref{sec:holothermo}.

%%--------------------------------------------------------------------
\subsection{Mass, temperature, entropy and the scaling isometry}
\label{sec:bulkthermo}
%%--------------------------------------------------------------------

The mass density $M$, the Hawking temperature $T$, and the entropy density $S$ are given by
\bea
&& M=\frac{m}{4\pi G_4}\,,\qquad
   T = \frac{3}{4\pi l^2}(2ml^2)^{1/3}e^{-\alpha/2}\,,\nonumber \\
&& S =\frac{(2ml^2)^{2/3}e^{\alpha/2}}{4G_4} \,.
\label{MTS}
\eea
At fixed $m$, a positive value of $\alpha$ enlarges $S$ and reduces $T$: the scalar field provides additional degrees of freedom and makes the system harder to excite, cooling it rather than heating it.
Our solution may therefore be called a \textit{scalar-cooled} AdS black brane.
Along curves of fixed $\alpha$ the first law takes the restricted form
\be
\label{firstlawfixedalpha}
\rmd S = \frac{4\pi l^2}{3}(8\pi G_4 M l^2)^{-1/3}e^{\alpha/2}\,\rmd M = \frac{\rmd M}{T}\,.
\ee

The structure of the solution space is governed by a scaling isometry.
Under
\be
\label{scaling}
r\to\lambda r\,,\qquad m\to\lambda^3 m\,,\qquad (t,x,y)\to\lambda^{-1}(t,x,y)\,,\qquad \alpha\ \text{fixed}\,,
\ee
the combination $\frac{\sinh\alpha}{ml^2}r^3$, and hence the scalar field, is invariant, and the metric (\ref{hairy}) is mapped to itself.
Solutions with the same $\alpha$ and different $m$ are thus related by a diffeomorphism combined with a rescaling of the boundary coordinates, under which $T\to\lambda T$, $S\to\lambda^2 S$ and $M\to\lambda^3 M$.
Accordingly, the thermodynamic quantities obey
\be
\label{EQSM}
S = \frac{1}{4 G_4}\bigg(\frac{4\pi l^2}{3}\bigg)^{\!2}e^{3\alpha/2}T^2\,,
\qquad
M = \frac{1}{6 G_4}\bigg(\frac{4\pi l^2}{3}\bigg)^{\!2}e^{3\alpha/2}T^3\,,
\ee
for all $(m,\alpha)$, together with the Smarr relation
\be
\label{Smarr}
3M=2TS\,,
\ee
which is the Smarr relation associated with the scaling (\ref{scaling}), in the same way that the Euler relation of holographic thermodynamics is dual to a generalized Smarr formula \cite{Karch:2015rpa,Visser:2021eqk,Mancilla:2024spp}.

%%--------------------------------------------------------------------
\subsection{Thermodynamic central charge and the generalized first law}
\label{sec:genfirstlaw}
%%--------------------------------------------------------------------

Taking the AdS/CFT correspondence into account and following the ABJM thermodynamics \cite{Bak:2010yd}, we define the thermodynamic central charge
\be
\label{Ct}
C_t = \frac{32l^2}{\pi G_4}\,e^{3\alpha/2}\,,
\ee
in terms of which (\ref{EQSM}) reads
\be
\label{SCt}
S=\frac{\pi^3l^2}{72}\,C_t\,T^2\,,\qquad M=\frac{\pi^3l^2}{108}\,C_t\,T^3\,.
\ee
$C_t$ roughly counts the effective number of degrees of freedom: it controls the overall normalization of the free energy in the same way as the ABJM central charge $\propto N^{3/2}$, and at $\alpha=0$ it reduces to $C_t^{(0)}\equiv32l^2/(\pi G_4)$.
It should not be identified with a conformal central charge in the strict sense, since the dual theory at $\alpha>0$ is not conformal.
Eliminating $T$ between the two relations in~(\ref{EQSM}) at fixed $\alpha$ gives
\be
\label{MSfixedalpha}
M = \frac{\sqrt{G_4}}{\pi l^2}\,e^{-3\alpha/4}\,S^{3/2}\,.
\ee
Writing $e^{3\alpha/2}\propto C_t$, the right-hand side becomes $S^{3/2}/C_t^{1/2}$, which is homogeneous of degree one in $(S,C_t)$ jointly, as in the holographic thermodynamics of CFTs, where the energy takes the form $C f(S/C)$ with a variable central charge $C$ \cite{Visser:2021eqk,Cong:2021fnf,Ahmed:2023snm}.
Treating $(S,C_t)$ as independent variables, the first law takes the generalized form
\bea
\label{genfirstlaw}
&& \rmd M = T\,\rmd S + \mu\,\rmd C_t\,,\qquad
   M(S,C_t) \equiv \frac{4\sqrt{2}\,S^{3/2}}{\pi^{3/2} l\,C_t^{1/2}}\,,
   \nonumber \\
&& T(S,C_t) \equiv \frac{6\sqrt{2}\,S^{1/2}}{\pi^{3/2} l\,C_t^{1/2}}\,,\qquad
   \mu(S,C_t) \equiv -\frac{2\sqrt{2}\,S^{3/2}}{\pi^{3/2} l\,C_t^{3/2}}\,,
\eea
valid on the whole two-parameter space of solutions, where $\mu$ is the chemical potential conjugate to $C_t$, and the Euler relation
\be
\label{Euler}
M = TS + \mu\, C_t
\ee
holds identically.
Combined with the Smarr relation (\ref{Smarr}), it gives $\mu C_t=-M/2$, so that $\mu<0$: at fixed entropy, increasing the scalar hair lowers the mass.

The Euler relation can equivalently be written in the fluid-like form $M = TS - Pv$, with $v\equiv \lambda\,C_t$ and $P\equiv -\mu/\lambda$ for an arbitrary length scale $\lambda$; here $P$ is a formal pressure conjugate to $v$, not the cosmological-constant pressure $P_\Lambda=3/(8\pi G_4 l^2)$ of black hole chemistry, and $v$ is a one-dimensional thermodynamic volume rather than the physical spatial volume.
For every $\lambda$ one finds $Pv = -\mu C_t = M/2$ and $\rmd M = T\,\rmd S - P\,\rmd v$.
Requiring that $v$ track the scalar hair alone and that $\lambda$ be built from the fixed data of the action, whose only dimensionful scale -- beyond the dimensionless ratio $l^2/G_4\sim N^{3/2}$ in the ABJM dual -- is the AdS radius $l$, selects
\be\label{vCt}
v \equiv l\,C_t\,,\qquad
P \equiv -\frac{\mu}{l} = \frac{2\sqrt{2}\,S^{3/2}}{\pi^{3/2} l^2\, C_t^{3/2}} > 0\,,
\ee
so that the hair parameter, encoded in $C_t\propto e^{3\alpha/2}$, plays the role of a transverse thermodynamic volume of the black brane.

The relations (\ref{EQSM})--(\ref{vCt}) characterize the black brane thermodynamics on the space of solutions.
In the holographic thermodynamics of CFTs \cite{Visser:2021eqk,Cong:2021fnf,Ahmed:2023snm,Gong:2023hgj} and in its restricted phase space formulation \cite{Zeyuan:2021uol}, the central charge is varied by changing $N$, i.e.\ by moving between different boundary theories.
For the present solution, varying $C_t$ means varying $\alpha$, which also changes the source of the dual operator.
To formulate the thermodynamics of a given dual theory we therefore need the boundary data and the one-point functions, which we now determine.

%%%%%%%%%%%%%%%%%%%%%%%%%%%%%%%%%%%%%%%%%%%%%%%%
\section{Holographic renormalization}
\label{sec:holoren}
%%%%%%%%%%%%%%%%%%%%%%%%%%%%%%%%%%%%%%%%%%%%%%%%

%%--------------------------------------------------------------------
\subsection{Asymptotic data}
\label{sec:asym}
%%--------------------------------------------------------------------

Near the asymptotic boundary the bulk scalar field has the expansion
\bea
\label{phiexpanded}
\phi_+(r) &=& \frac{\sqrt{7}}{2}\ln\frac{l\sinh\alpha}{m}
              + \frac{3\sqrt{7}}{2}\ln\!\left(\frac{r}{l}\right)
              + \frac{\sqrt{7}\,me^{-\alpha}}{2l\sinh\alpha}
                \!\left(\frac{l}{r}\right)^{\!3} + \cdots
              \nonumber \\
          &\equiv& \varphi_{(0)} + \varphi_{(1)}\ln\!\left(\frac{r}{l}\right)
                  + \varphi_{(3)}\!\left(\frac{l}{r}\right)^{\!3} + \cdots\,,
\eea
with
\be
\label{phi0phi3}
\varphi_{(0)}=-\frac{3\sqrt7}{2}\ln\frac{r_s}{l}\,,\qquad
\varphi_{(1)}=\frac{3\sqrt7}{2}\,,\qquad
\varphi_{(3)}=\frac{\sqrt7}{2}\,e^{-\alpha}\Big(\frac{r_s}{l}\Big)^{3}\,.
\ee
The coefficient $\varphi_{(1)}$ of the logarithm is not a free parameter: it is fixed by the exponent of the leading term $e^{-3\phi_+/\sqrt7}$ of the potential, as follows for instance from the domain-wall flow $\rmd\phi_+/\rmd A=-4W_-'/W_-\to6/\sqrt7$ with $e^{A}\propto r^{7/4}$, and it takes the same value for the whole class of solutions of \cite{Cadoni:2011sm}.
The free asymptotic data of the scalar are therefore $\varphi_{(0)}$ and $\varphi_{(3)}$.
The non-normalizable part $\varphi_{(0)}+\varphi_{(1)}\ln(r/l)$ corresponds to the source of the dual operator $\mathcal{O}_\phi$, representing the UV coupling strength, with $\varphi_{(1)}$ characterizing its logarithmic running with the energy scale.
The source depends on the solution only through $r_s$, whereas $\varphi_{(3)}$ carries the dependence on the state.
With the solution-independent defining function $(r/l)^{7/2}$, the induced metric on a surface of constant $r$ behaves as
\be
\label{bdymetric}
\gamma_{ij}\,\rmd x^i\rmd x^j\simeq\Big(\frac{r}{l}\Big)^{7/2}\hat g_{(0)ij}\,\rmd x^i\rmd x^j\,,\qquad
\hat g_{(0)}=\Big(\frac{l}{r_s}\Big)^{3/2}\Big(-\rmd t^2+l^2\rmd x^2+l^2\rmd y^2\Big)\,,
\ee
i.e.\ $\sqrt{g_{xx}}\simeq r_s^{-3/4}\,r^{7/4}$.
The boundary data of the dual theory, the flat metric $\hat g_{(0)}$ and the source $\varphi_{(0)}$, thus depend on the solution only through $r_s$.
In the boundary coordinates
\be
\label{bdycoords}
(\hat t,\hat x,\hat y)=\Big(\frac{l}{r_s}\Big)^{3/4}(t,x,y)
\ee
the boundary metric takes the fixed form $\hat g_{(0)}=-\rmd\hat t^2+l^2\rmd\hat x^2+l^2\rmd\hat y^2$.

%%--------------------------------------------------------------------
\subsection{Counterterms and one-point functions}
\label{sec:onept}
%%--------------------------------------------------------------------

Because the UV asymptotics is that of a non-conformal brane rather than AdS, we renormalize the on-shell action following the approach developed for non-conformal branes \cite{Kanitscheider:2008kd} and the Hamilton--Jacobi formulation of the holographic renormalization group \cite{Boer:2000}, which generalize the AdS procedure \cite{deHaro:2000vlm,Balasubramanian:1999re,Bianchi:2001kw}.
On a cutoff surface of constant $r$ with induced metric $\gamma_{ij}$, outward unit normal $n$ and extrinsic curvature $K_{ij}=\frac12 n^r\partial_r\gamma_{ij}$, the renormalized action is
\be
\label{Iren}
I_{\rm ren}=I+\frac{1}{8\pi G_4}\int\rmd^3x\sqrt{-\gamma}\,K-\frac{1}{8\pi G_4}\int\rmd^3x\sqrt{-\gamma}\;\frac{W(\phi_+)}{2}\,,
\ee
where $W$ solves the Hamilton--Jacobi equation $V=\frac12[(W')^2-\frac34W^2]$ for flat slices with the leading asymptotics $W\simeq\frac7l e^{-3\phi_+/2\sqrt7}$ \cite{Boer:2000,Kanitscheider:2008kd}.
For a flat boundary no curvature counterterms are needed.
The two exact solutions $W_\pm$ of eq.~(\ref{fakeW}) differ by $\frac6l e^{-\sqrt7\phi_+/2}$, and $\sqrt{-\gamma}\,e^{-\sqrt7\phi_+/2}\to r_s^3/l$ is finite at the boundary; the coefficient of this term is therefore a finite counterterm, i.e.\ a choice of scheme.
We adopt the scheme $W=W_-$, defined by the superpotential of the zero-temperature domain wall, in which the vacuum has vanishing one-point functions.

We first evaluate the one-point functions as densities per unit coordinate area in the coordinates $(t,x,y)$ of (\ref{hairy}), with energies conjugate to $t$, i.e.\ in the same units as $M$, $T$ and $S$ in (\ref{MTS}); the conversion to the boundary coordinates (\ref{bdycoords}) is made in Section~\ref{sec:holothermo}.
With the Brown--York tensor $T_{ij}[\gamma]=\frac{1}{8\pi G_4}\big(-K_{ij}+K\gamma_{ij}-\frac12W_-\gamma_{ij}\big)$, the energy density, the pressure and the scalar one-point function are
\bea
\label{onept_def}
&&\varepsilon=-\lim_{r\to\infty}\sqrt{-\gamma}\,T^t{}_t[\gamma]\,,\qquad
p=\lim_{r\to\infty}\sqrt{-\gamma}\,T^x{}_x[\gamma]=\lim_{r\to\infty}\sqrt{-\gamma}\,T^y{}_y[\gamma]\,,\nonumber\\
&&\langle\mathcal{O}_\phi\rangle=\frac{\delta I_{\rm ren}}{\delta\varphi_{(0)}}=-\lim_{r\to\infty}\frac{\sqrt{-\gamma}}{16\pi G_4}\Big(n^r\partial_r\phi_++W_-'(\phi_+)\Big)\,,
\eea
where the variation is taken at fixed $\hat g_{(0)}$.
All divergences cancel, as shown in Appendix~\ref{app:holoren}, and for all $(m,\alpha)$ we find
\bea
\label{holoresults}
&&\varepsilon=\Big(1+\frac34\tanh\frac{\alpha}{2}\Big)M\,,\qquad
p=\Big(\frac12-\frac34\tanh\frac{\alpha}{2}\Big)M\,,\nonumber\\
&&\langle\mathcal{O}_\phi\rangle=\frac{3\sqrt7}{8}\,M\tanh\frac{\alpha}{2}\,.
\eea
The energy density can therefore be written as
\be
\label{epsM}
\varepsilon=M+\frac{2}{\sqrt7}\,\langle\mathcal{O}_\phi\rangle\,:
\ee
the scalar condensate contributes to the energy density, and $\varepsilon$ and $M$ coincide only for $\alpha\to0$.
In terms of the asymptotic coefficients,
\be
\label{Ophi}
\langle\mathcal{O}_\phi\rangle=\frac{3}{16\pi G_4}\Big(\frac{\sqrt7}{2}\,m+l\,\varphi_{(3)}\Big)-\frac{3\sqrt7}{32\pi G_4}\,\frac{r_s^3}{l^2}\,,
\ee
so that, as is typical for non-AdS asymptotics, the one-point function mixes the scalar coefficient $\varphi_{(3)}$ with the normalizable coefficient $m$ of the metric.
In particular, $\langle\mathcal{O}_\phi\rangle$ is not proportional to $\varphi_{(3)}$, and it vanishes in the limit $\alpha\to0$ even though $\varphi_{(0)}$ diverges there at fixed $m$.

In a general scheme, $W_-\to W_-+\frac{c+3}{l}e^{-\sqrt7\phi_+/2}$, the one-point functions shift by constants,
\bea
\label{scheme}
&&\varepsilon\to\varepsilon+(c+3)k\,,\qquad p\to p-(c+3)k\,,\nonumber\\
&&\langle\mathcal{O}_\phi\rangle\to\langle\mathcal{O}_\phi\rangle+\frac{\sqrt7}{2}(c+3)k\,,\qquad
k\equiv\frac{M}{4\sinh\alpha}=\frac{r_s^3}{16\pi G_4 l^2}\,,
\eea
which depend on the source only and leave all $\alpha$-dependent quantities unchanged.

%%--------------------------------------------------------------------
\subsection{Ward identity}
\label{sec:ward}
%%--------------------------------------------------------------------

The trace of the stress tensor follows from (\ref{holoresults}),
\be
\label{trace}
\langle T^i{}_i\rangle=-\varepsilon+2p=-\frac94\,M\tanh\frac{\alpha}{2}\,,
\ee
and satisfies the Ward identity
\be
\label{Ward}
\langle T^i{}_i\rangle=-\beta_\infty\,\langle\mathcal{O}_\phi\rangle\,,\qquad \beta_\infty=\frac{4\varphi_{(1)}}{7}=\frac{6\sqrt7}{7}\,,
\ee
where $\beta_\infty$ coincides with the UV limit of the holographic beta function (\ref{betafunction}).
This identity follows from the invariance of $I_{\rm ren}$ under the radial reparametrization $r\to\lambda r$, which by (\ref{phiexpanded}) and (\ref{bdymetric}) acts on the boundary data as the Weyl rescaling $\hat g_{(0)}\to\lambda^{7/2}\hat g_{(0)}$ accompanied by the shift $\varphi_{(0)}\to\varphi_{(0)}+\varphi_{(1)}\ln\lambda$; since the counterterms in (\ref{Iren}) are covariant, this gives $\frac74\langle T^i{}_i\rangle+\varphi_{(1)}\langle\mathcal{O}_\phi\rangle=0$ \cite{Bianchi:2001kw,Kanitscheider:2008kd}.
The identity holds in any scheme (\ref{scheme}).
Hence the stress tensor of the deformed theory is traceless only in the limit $\alpha\to0$, where the condensate vanishes.
In Section~\ref{sec:holofirstlaw} we show that (\ref{Ward}) is also the scaling relation of the holographic free energy and the boundary image of the bulk Smarr relation (\ref{Smarr}).

%%%%%%%%%%%%%%%%%%%%%%%%%%%%%%%%%%%%%%%%%%%%%%%%
\subsection{Status of the holographic dictionary}
\label{sec:dictionary}

The general results of holographic renormalization \cite{deHaro:2000vlm,Bianchi:2001kw} rely on a Fefferman--Graham expansion about an asymptotically AdS metric and do not cover the asymptotics (\ref{UVmetric}).
The Hamilton--Jacobi method \cite{Boer:2000,Kanitscheider:2008kd} used above does apply, and we adopt the resulting dictionary as a set of explicit working assumptions:
(A1) the renormalized on-shell action and the one-point functions are those obtained above with the counterterm $W_-$;
(A2) $\varphi_{(0)}$ is the source and $\langle\mathcal O_\phi\rangle$ the response, an identification we use only for static configurations;
(A3) where the lower bound of the energy is discussed (Section~\ref{sec:multitrace}), the criterion of Hertog and Horowitz \cite{Hertog:2004ns}, formulated for asymptotically AdS spacetimes, applies.
The evidence for (A1) and (A2) is internal consistency: the divergences cancel, and the one-point functions satisfy the holographic first law, the Ward identity and the Euler relation identically on the two-parameter space of solutions (Section~\ref{sec:holothermo}).
In \cite{Cadoni:2012uf} a holographic interpretation was restricted to the range of potentials for which the asymptotic region is a timelike boundary, $\beta^2<1/3$ in the notation used there; the present model, with $\beta^2=3/7$, lies outside this range, which is why the dictionary is adopted here as a set of assumptions rather than derived.

The two asymptotic data of the scalar differ in their normalizability.
The Klein--Gordon norm of a scalar perturbation involves $\int\rmd r\,\sqrt{-g}\,|g^{tt}|\,|\delta\phi_+|^2$, and on (\ref{hairy})
\be
\label{KGweight}
\sqrt{-g}\,|g^{tt}|=\frac{r^2\Big[\big(\frac{r}{r_s}\big)^3+e^{-\alpha}\Big]}{\frac{r^2}{l^2}-\frac{2m}{r}}\simeq\frac{l^2r^3}{r_s^3}\,.
\ee
The state-dependent mode $\delta\phi_+\propto r^{-3}$ is therefore normalizable, while the source mode $\delta\phi_+=\delta\varphi_{(0)}$ is not, its norm diverging as $r^4$.
This is the analogue of the source mode of a marginal operator in AdS, which is not normalizable either \cite{Breitenlohner:1982jf,Klebanov:1999tb}.
The source is therefore a coupling of the dual theory, and holding it fixed, as in the Dirichlet theory of Section~\ref{sec:holothermo}, is the standard choice.
For time-dependent perturbations the asymptotic region at infinite light-crossing time admits no source/condensate split at all (Appendix~\ref{app:dynamics}), so that transport coefficients are extracted below from static horizon data.

%%%%%%%%%%%%%%%%%%%%%%%%%%%%%%%%%%%%%%%%%%%%%%%%%%%%%%%%%%%%%%%%%%
\section{Holographic thermodynamics}
\label{sec:holothermo}
%%%%%%%%%%%%%%%%%%%%%%%%%%%%%%%%%%%%%%%%%%%%%%%%

%%--------------------------------------------------------------------
\subsection{Variables and the holographic first law}
\label{sec:holofirstlaw}
%%--------------------------------------------------------------------

The dual field theory is specified by the boundary metric $\hat g_{(0)}$ and the source $\varphi_{(0)}$.
We refer to the theory in which $\varphi_{(0)}$ is held fixed as the Dirichlet theory; an alternative boundary condition is discussed in Section~\ref{sec:multitrace}.
When the source is varied, the boundary metric must be held fixed, and boundary quantities must therefore be measured in the coordinates (\ref{bdycoords}).
We denote the densities in these coordinates by a hat; they are related to those of Sections~\ref{sec:thermo} and~\ref{sec:onept} by
\be
\label{hatconv}
\hat T=\Big(\frac{r_s}{l}\Big)^{3/4}T\,,\qquad \hat s=\Big(\frac{r_s}{l}\Big)^{3/2}S\,,\qquad
\big(\hat\varepsilon,\hat p,\hat{\mathcal{O}}\big)=\Big(\frac{r_s}{l}\Big)^{9/4}\big(\varepsilon,p,\langle\mathcal{O}_\phi\rangle\big)\,.
\ee
For a fixed boundary theory these factors are constants, and dimensionless ratios such as $\hat s/\hat T^2=S/T^2$ or $\hat\varepsilon/\hat T^3=\varepsilon/T^3$ are unaffected.

The invariance of the generating functional under the transformation of Section~\ref{sec:ward}, a constant Weyl rescaling of $\hat g_{(0)}$ combined with a shift of $\varphi_{(0)}$, implies that the theories with different sources are related by a change of units.
The source therefore defines a single, renormalization-group invariant scale,
\be
\label{Lambda}
\Lambda\equiv\frac1l\,e^{-\varphi_{(0)}/\beta_\infty}=\frac1l\Big(\frac{r_s}{l}\Big)^{7/4}\,,
\ee
which plays a role analogous to $\Lambda_{\rm QCD}$, and every thermodynamic density is $\Lambda$ to the appropriate power times a function of $\hat T/\Lambda$.
From (\ref{MTS}), (\ref{rs}) and (\ref{hatconv}),
\be
\label{TLambda}
\frac{\hat T}{\Lambda}=\frac{l^2T}{r_s}=\frac{3}{4\pi}\,(2\sinh\alpha)^{1/3}e^{-\alpha/2}\,,\qquad \Big(\frac{r_h}{r_s}\Big)^3=2\sinh\alpha\,,
\ee
so that, in a given boundary theory, $\alpha$ is not an independent variable but determines the temperature in units of $\Lambda$.
The scaling isometry (\ref{scaling}), which changes $r_s$ at fixed $\alpha$, rescales $\hat T$ and $\Lambda$ together and amounts to a change of units.

The thermodynamic potential of the boundary theory is the free energy density $\hat f(\hat T,\varphi_{(0)})$, obtained from the renormalized Euclidean on-shell action.
Its natural variables are the temperature and the source, or equivalently $\hat T$ and $\Lambda$, and since $\langle\mathcal{O}_\phi\rangle$ is the response to the source,
\be
\label{dfree}
\rmd\hat f=-\hat s\,\rmd\hat T-\hat{\mathcal{O}}\,\rmd\varphi_{(0)}\,,\qquad \hat f=-\hat p\,.
\ee
With the Euler relation
\be
\label{holoEuler}
\hat\varepsilon+\hat p=\hat T\hat s\,,
\ee
which expresses the extensivity of the planar system in the transverse directions, this gives the holographic first law
\be
\label{holofirstlaw}
\rmd\hat\varepsilon=\hat T\,\rmd\hat s-\hat{\mathcal{O}}\,\rmd\varphi_{(0)}=\hat T\,\rmd\hat s+\beta_\infty\hat{\mathcal{O}}\,\frac{\rmd\Lambda}{\Lambda}\,,
\ee
in which the variables are the entropy density and the source, and the work term is that of the source.
Since $\hat f$ has dimension three, $\hat f=\Lambda^3\mathfrak{f}(\hat T/\Lambda)$, and the scaling relation $\hat T\,\partial_{\hat T}\hat f+\Lambda\,\partial_\Lambda\hat f=3\hat f$ combined with (\ref{dfree})--(\ref{holoEuler}) yields
\be
\label{holoscaling}
\hat\varepsilon-2\hat p=\beta_\infty\,\hat{\mathcal{O}}\,,
\ee
i.e.\ the Ward identity (\ref{Ward}).
The conjugate of $\ln\Lambda$ is thus $\beta_\infty\hat{\mathcal{O}}=-\langle\hat T^i{}_i\rangle$: the trace of the stress tensor measures the response of the free energy to the intrinsic scale.
We have verified that the one-point functions (\ref{holoresults}), converted by (\ref{hatconv}), satisfy (\ref{dfree})--(\ref{holoscaling}) identically on the full two-parameter space of solutions, and that the renormalized Euclidean on-shell action reproduces $\hat f=-\hat p$ (Appendix~\ref{app:holoren}).

The black hole thermodynamics of Section~\ref{sec:thermo} and the holographic thermodynamics are two descriptions of the same solutions, and the relations between them are instructive.
The bulk description uses the variables $(S,C_t)$ with conjugates $(T,\mu)$, while the boundary description uses $(\hat s,\varphi_{(0)})$ with conjugates $(\hat T,-\hat{\mathcal{O}})$.
The bulk Smarr relation (\ref{Smarr}) and the boundary scaling relation (\ref{holoscaling}) express the same scaling isometry: using (\ref{epsM}) and $p=TS-\varepsilon$, one finds $\varepsilon-2p=3\varepsilon-2TS=3M-2TS+\frac{6}{\sqrt7}\langle\mathcal{O}_\phi\rangle$, so that (\ref{Smarr}) implies (\ref{holoscaling}).
Along the thermal states of a fixed theory, $\rmd\varphi_{(0)}=0$ while $C_t$ varies, and comparing (\ref{genfirstlaw}) with (\ref{holofirstlaw}),
\be
\label{muOrelation}
\mu\,\rmd C_t\big|_{\varphi_{(0)}}=\rmd M-T\,\rmd S\big|_{\varphi_{(0)}}=-\frac{2}{\sqrt7}\,\rmd\langle\mathcal{O}_\phi\rangle\big|_{\varphi_{(0)}}\,:
\ee
the work term of the thermodynamic central charge in the bulk first law is precisely the change in the condensate contribution to the boundary energy density.
Finally, the thermodynamic central charge becomes a derived quantity of the boundary theory,
\be
\label{Ctderived}
C_t=\frac{72}{\pi^3l^2}\,\frac{\hat s}{\hat T^2}\,,
\ee
a function of $\hat T/\Lambda$ that counts the effective number of degrees of freedom at temperature $\hat T$, in the same way as $s/T^3$ does in four-dimensional gauge theories.

%%--------------------------------------------------------------------
\subsection{Thermal states and phase structure}
\label{sec:fixedtheory}
%%--------------------------------------------------------------------

We now consider the thermal states of a fixed boundary theory, i.e.\ fixed $\Lambda$, and omit the hats, since only ratios to powers of $\Lambda$ matter.
By (\ref{TLambda}), the temperature is not a monotonic function of $\alpha$: it vanishes both as $\alpha\to0$ and as $\alpha\to\infty$ and reaches the maximum
\be
\label{Tmax}
T_{\max}=\frac{3}{4\pi}\,2^{2/3}5^{-5/12}\,\Lambda\simeq0.194\,\Lambda\qquad\text{at}\qquad \alpha_*=\tfrac12\ln5\simeq0.805\,.
\ee
Below $T_{\max}$ there are two black branes at each temperature, and above it there is none in the present family.
The specific heat at fixed $\Lambda$,
\be
\label{heatcap}
C_\Lambda\equiv T\frac{\partial s}{\partial T}\Big|_\Lambda=s\,\frac{4\coth\alpha+3}{2\coth\alpha-3}\,,
\ee
is positive on the lower branch $0<\alpha<\alpha_*$ and negative on the upper branch $\alpha>\alpha_*$.

The two branches probe different regions of the geometry.
On the lower branch, $T\to0$ as $\alpha\to0$ and $s\to\frac{1}{4G_4}\big(\frac{4\pi l^2}{3}\big)^2T^2=\frac{\pi^3l^2}{72}C_t^{(0)}T^2$: the horizon lies in the region $r\ll r_s$, where the geometry is close to AdS$_4$ with $\phi_+\simeq0$, and the entropy approaches that of the ABJM plasma.
On the upper branch, $s\propto\Lambda^9T^{-7}$ as $\alpha\to\infty$, i.e.\ $s\propto T^{(d-\theta)/z}$ with the hyperscaling-violating exponents $\theta=9$ and $z=1$ of the UV region (Section~\ref{sec:rg}) \cite{Charmousis:2010zz,Dong:2012se}; the negative specific heat is the familiar thermodynamic instability of hyperscaling-violating black branes with $\theta>d$ \cite{Charmousis:2010zz,Gouteraux:2011ce}.
Accordingly, the thermodynamic central charge (\ref{Ctderived}) runs with temperature, from the ABJM value $C_t^{(0)}$ at $T\to0$ on the lower branch to $C_t\propto(\Lambda/T)^9$ on the upper branch.

Along the thermal path the first law (\ref{holofirstlaw}) reduces to $\rmd\varepsilon=T\,\rmd s$, while the bulk mass density satisfies $\rmd M-T\,\rmd S=\mu\,\rmd C_t\neq0$ by (\ref{muOrelation}).
The free energy density relative to the zero-temperature domain wall is
\be
\label{freeenergy}
f=-p=\frac{l^4\Lambda^3}{16\pi G_4}\Big[3(\cosh\alpha-1)-2\sinh\alpha\Big]\,,
\ee
which is independent of the scheme (\ref{scheme}), since the constant shifts cancel against those of the vacuum.
On the lower branch $f<0$, so that the black brane is thermodynamically preferred over the domain wall at every temperature $0<T<T_{\max}$.
This is consistent with the absence of a Hawking--Page transition for planar horizons, and the boundary system is gapless and deconfined there.
At a given temperature the upper branch has a higher free energy, which becomes positive for $\alpha>\ln5$; together with its negative specific heat, this makes the upper branch thermodynamically disfavored.
The existence of a maximal temperature, the negative specific heat and the negative $c_s^2$ found below on the upper branch arise from the UV exponential $e^{-3\phi_+/\sqrt7}$, while the second exponential in (\ref{action}) provides the AdS$_4$ fixed point that stabilizes the lower branch.

These results should be compared with those of \cite{Cadoni:2012uf}, whose one-parameter family of solutions is, for $\beta^2=3/7$, the set of thermal states of the Dirichlet theory.
For this value of $\beta$ the formulas given there yield $\theta=9$ and $z=1$, a maximal temperature attained at $\nu_2=1/\sqrt5$, i.e.\ at $\alpha_*$ of (\ref{Tmax}), and two branches of black branes below it, in agreement with the above.
Moreover, up to the normalization of the time coordinate, the mass and the free energy computed there differ from our $\varepsilon$ and $f$ only by the additive constant $3\,l^4\Lambda^3/(16\pi G_4)$, which is the shift (\ref{scheme}) with $c=0$.
The conclusion drawn in \cite{Cadoni:2012uf} is nevertheless that both branches are unstable, because the free energy so normalized is positive whereas that of the AdS-Schwarzschild brane is negative.
The difference from our conclusion has two origins.
First, the additive constant is the free energy assigned to the zero-temperature domain wall; it depends on the source only and is a finite counterterm, so that the sign of the free energy has no scheme-independent meaning, whereas the difference (\ref{freeenergy}) from the domain wall does.
Second, the AdS-Schwarzschild brane has $\phi_+=0$ and corresponds to $\Lambda\to\infty$ (Section~\ref{sec:hydro}); it is not a state of the theory with a given finite $\Lambda$ and does not compete with the hairy black branes in the canonical ensemble at fixed source.
The first-order transition between the AdS-Schwarzschild and the scalar-dressed branes found in \cite{Cadoni:2012uf} occurs for $\beta^2<1/3$, where $\theta<0$, and has no counterpart in the present model.

%%%%%%%%%%%%%%%%%%%%%%%%%%%%%%%%%%%%%%%%%%%%%%%%
\section{Holographic RG flow and hydrodynamics}
\label{sec:boundary}
%%%%%%%%%%%%%%%%%%%%%%%%%%%%%%%%%%%%%%%%%%%%%%%%

%%--------------------------------------------------------------------
\subsection{Holographic RG flow}
\label{sec:rg}
%%--------------------------------------------------------------------

For $\alpha=0$, our solution reduces to pure AdS-Schwarzschild, whose dual boundary theory is a three-dimensional CFT at finite temperature with exact conformal symmetry.
For positive $\alpha$, the large-$r$ region of the hairy black brane is an asymptotically conformally AdS geometry with hyperscaling violation exponent $\theta=9$ and dynamical critical exponent $z=1$,\footnote{We use the standard convention of \cite{Gouteraux:2011ce,Huijse:2011ef,Dong:2012se}, in which the metric $ds^2_{d+2}=r^{-2(d-\theta)/d}\bigl(-r^{-2(z-1)}dt^2+dr^2+dx_i^2\bigr)$ transforms as $ds\to\lambda^{\theta/d}ds$ under $x_i\to\lambda x_i$, $t\to\lambda^z t$, $r\to\lambda r$, with $d$ the number of boundary spatial dimensions ($d=2$ here).} obtained by matching the large-$r$ asymptotics of (\ref{hairy}) to the standard hyperscaling-violating form; the corresponding geometry satisfies the null energy conditions for these exponents.
Domain walls supported by exponential potentials are among the few non-AdS geometries with a precise holographic formulation, through a generalized conformal structure \cite{Kanitscheider:2008kd,Lima:2020}.
In the zero-temperature domain wall, on the other hand, the region $r\ll r_s$ approaches AdS$_4$ with $\phi_+=0$.
The boundary theory is therefore a three-dimensional QFT whose UV regime is controlled by an operator of effective dimension $\Delta_{\rm eff}\to 3^-$ with a logarithmically running source -- a marginally relevant deformation in this sense \cite{Park:2021} -- and which flows to the ABJM fixed point in the IR.
In the UV regime it is described schematically by the effective action
\be
S_{\rm bdy} = S_{\rm CFT}
  + \int \rmd^3x\,\bigg[\varphi_{(0)}
    + \varphi_{(1)}\ln\!\bigg(\frac{\mu_{\Lambda}}{l}\bigg)\bigg]\mathcal{O}_\phi
  + \cdots\,,
\ee
where $\mu_{\Lambda}$ is the renormalization scale and the dots denote further operators generated along the flow; this description applies where the effective dimension of $\mathcal{O}_\phi$ is close to three, and not near the IR fixed point.
The effective dimension follows from $m^2_{\rm eff}=V''(\phi_+)$, which vanishes as $r\to\infty$ and is approached from below, since $m^2_{\rm eff} < 0$ whenever $\phi_+>\frac{\sqrt7}{4}\ln(7/3)\simeq0.560$, i.e.\ throughout the scalar-dominated region.
Near $\phi_+=0$, by contrast, $m^2_{\rm eff}\simeq18/l^2>0$, so that $\mathcal{O}_\phi$ is irrelevant with $\Delta=6$ at the IR fixed point.
The logarithmic deformation reduces the three-dimensional conformal symmetry $SO(3,2)$ to the Poincar\'{e} group $ISO(2,1)$, and the full symmetry is recovered at the IR fixed point.

Explicitly, $m^2_{\rm eff} = V''(\phi_+) = -\frac{27}{2l^2}e^{-3\phi_+/\sqrt{7}} +\frac{63}{2l^2}e^{-\sqrt{7}\phi_+}$, with minimum $m^2_{\rm min}\approx -2.164/l^2$ at $\phi_+ = \frac{\sqrt{7}}{2}\ln(7/3)\approx 1.121$, which lies above the Breitenlohner--Freedman bound $-9/(4l^2)$ \cite{Breitenlohner:1982jf}, as required for the stability of the holographic RG flow.
The effective conformal dimension $\Delta_{\rm eff} = \frac{3}{2}+\sqrt{\frac{9}{4}+m^2_{\rm eff}l^2}$ has minimum $\Delta_{\rm min}\approx 1.7926$, above the unitarity bound.
Notably, $V''(\phi_+(r_h)) = 0$ when $\alpha = \frac{1}{2}\ln(7/3)$ or $\alpha\to\infty$, so that the effective dimension evaluated at the horizon also reaches the marginal value $\Delta_{\rm eff}(r_h) = 3$.

The holographic beta function \cite{Boer:2000}, computed from $\beta = \rmd\phi_+/\rmd A$ with $e^A = (r/l)(\frac{\sinh\alpha}{ml^2}r^3+e^{-\alpha})^{1/4}$, is
\be
\label{betafunction}
\beta(\phi_+) = \frac{6\sqrt{7}}{7+\frac{4me^{-\alpha}}{l\sinh\alpha}
                \big(\frac{l}{r}\big)^3}
= \frac{6\sqrt{7}}{7 + \frac{4}{e^{\alpha+2\phi_+/\sqrt{7}}-1}}\,.
\ee
In the UV limit this approaches the constant $\beta_\infty=6\sqrt{7}/7$, confirming logarithmic running; this is the coefficient appearing in the Ward identity (\ref{Ward}).

The beta function also controls the c-function of Section~\ref{sec:solution}.
With the same warp factor $e^A=h(r)/l$, a direct computation gives, for all $(m,\alpha)$,
\be
\label{cbeta}
\frac{\rmd\ln c}{\rmd A}=\frac12\,\beta(\phi_+)^2\,,
\ee
so that $c$ increases monotonically towards the UV precisely because the coupling runs, and is stationary only where $\beta=0$.
Along the zero-temperature domain wall, $\beta\to0$ and $c\to1$, the AdS$_4$ value, at the ABJM fixed point, where $\mathcal{O}_\phi$ has $\Delta=6$; in the UV, $\beta\to\beta_\infty$ and $c\propto e^{\beta_\infty^2A/2}=e^{18A/7}$ grows without bound, as expected for the hyperscaling-violating regime, where $\Delta_{\rm eff}\to3^-$.
Eq.~(\ref{cbeta}) is the analogue, for the present exact flow, of the holographic c-theorems of \cite{Freedman:1999gp,Astefanesei:2007vh}, and it ties the monotonicity of $c$ directly to the running of the source.

%%--------------------------------------------------------------------
\subsection{Hydrodynamics and the emergent IR conformal fluid}
\label{sec:hydro}
%%--------------------------------------------------------------------

For a fluid without conserved charges the speed of sound squared is $c_s^2=\rmd p/\rmd\varepsilon$ along the thermal states of a fixed theory.
From (\ref{holoresults}) and (\ref{TLambda}),
\be
\label{cs2}
c_s^2=\frac{2\coth\alpha-3}{4\coth\alpha+3}\,.
\ee
This result admits an independent cross-check that uses only the horizon data: for a single-component fluid satisfying $\rmd\varepsilon=T\,\rmd s$, the speed of sound obeys the caloric identity $c_s^2 = \rmd(\ln T)/\rmd(\ln s)$, which applied to (\ref{MTS}) at fixed $r_s$ reproduces (\ref{cs2}).
On the lower branch $c_s^2$ decreases from the conformal value $1/2$ \cite{Baier:2007dw} at $T\to0$ and vanishes at $T_{\max}$, where $\rmd T/\rmd s=0$; on the upper branch $c_s^2<0$ and tends to $-1/7=z/(d-\theta)$, the value dictated by hyperscaling violation.

The bulk viscosity follows from the Eling--Oz formula \cite{Eling:2011ms}, which in the normalization of (\ref{action}) reads $\zeta/\eta=(s\,\rmd\phi_+(r_h)/\rmd s)^2$, with the derivative taken at fixed $\Lambda$.
Using $\phi_+(r_h)=\frac{\sqrt7}{2}\alpha$,
\be
\label{zeta}
\frac{\zeta}{\eta}=\frac{63\sinh^2\alpha}{(3\sinh\alpha+4\cosh\alpha)^2}\,.
\ee
The bulk viscosity vanishes only in the IR limit $\alpha\to0$, where the horizon value of the scalar becomes independent of the entropy, and $\zeta/\eta\to9/7$ as $\alpha\to\infty$.
As a consistency check of the latter value, the relation $\zeta/\eta = 2(1/(D_{bdy}-1) - c_s^2)$ of \cite{Kanitscheider:2009mn}, with $D_{bdy}=3$ the boundary \emph{spacetime} dimension, gives $\zeta/\eta=9/7$ for $c_s^2=-1/7$.
That relation was derived for theories with the generalized conformal structure of a single exponential potential, and we use it only in the UV regime $\alpha\to\infty$, where the potential is dominated by $e^{-3\phi_+/\sqrt7}$; at finite $\alpha$ it does not apply, and indeed differs from (\ref{zeta}).
At finite $\alpha$, (\ref{zeta}) lies below the bound $\zeta/\eta\ge2\big(1/(D_{bdy}-1)-c_s^2\big)$ conjectured by Buchel \cite{Buchel:2007mf}: with (\ref{cs2}) the right-hand side equals $9/(4\coth\alpha+3)$, and the ratio of the two sides is $7/(4\coth\alpha+3)<1$ for all $\alpha>0$.
The violation is strongest at low temperature, where $\frac12-c_s^2$ is of first order in $(T/\Lambda)^3$ while $\zeta/\eta$ is of second order, and the bound is approached only as $\alpha\to\infty$; violations of this bound have also been found in other holographic models \cite{Gubser:2008sz,Buchel:2011wx}.
The shear viscosity saturates the universal value $\eta/s=1/(4\pi)$ \cite{Policastro:2001yc,Kovtun:2004de} for all $\alpha$, as for any two-derivative Einstein-gravity dual.

The approach to the IR fixed point can be made explicit.
Writing $x\equiv\frac{4\pi T}{3\Lambda}=(2\sinh\alpha)^{1/3}e^{-\alpha/2}$ and inverting on the lower branch, $\alpha=\frac12x^3+\frac38x^6+O(x^9)$, we find
\bea
\label{lowT}
&&\varepsilon=\frac{\pi^3l^2}{108}\,C_t^{(0)}T^3\Big(1+\frac{15}{16}x^3+\cdots\Big)\,,\nonumber\\
&&p=\frac{\pi^3l^2}{216}\,C_t^{(0)}T^3\Big(1+\frac38x^3+\cdots\Big)\,,\nonumber\\
&&\frac{\langle T^i{}_i\rangle}{\varepsilon}=-\frac{9}{16}\,x^3+\cdots\,,\qquad
\frac{\langle\mathcal{O}_\phi\rangle}{\varepsilon}=\frac{3\sqrt7}{32}\,x^3+\cdots\,,\nonumber\\
&&c_s^2=\frac12-\frac{9}{16}x^3+\cdots\,,\qquad
\frac{\zeta}{\eta}=\frac{63}{64}\,x^6+\cdots\,.
\eea
The stress tensor therefore takes the form
\be\label{Tij_CT}
\langle T^i{}_j\rangle=\mathrm{diag}\big(-\varepsilon,\,p,\,p\big)
=\frac{\pi^3l^2}{108}\,C_t^{(0)}\,T^3\,\mathrm{diag}\!\bigl(-1,\tfrac{1}{2},\tfrac{1}{2}\bigr)+O\big(T^6/\Lambda^3\big)\,,
\ee
with $\varepsilon$ and $p$ given exactly by (\ref{holoresults}).
The leading term is the stress tensor of the ABJM plasma, with $C_t^{(0)}$ playing the role of the $N^{3/2}$ factor of ABJM theory \cite{Aharony:2008ug,Bak:2010yd}.
At the undeformed point the tracelessness $\langle T^i{}_i\rangle=0$ is guaranteed by the conformal Ward identity: three-dimensional CFTs have no Weyl anomaly, so $\langle T^i{}_i\rangle=0$ holds to all orders in the $1/N$ expansion.
In the deformed theory the trace is given by (\ref{Ward}) and is suppressed at low temperature.

The boundary theory thus exhibits a holographic RG flow from a non-conformal UV to the conformal ABJM fixed point, and its thermal state is a nearly conformal fluid for $T\ll\Lambda$: the horizon screens the UV deformation, and conformality, while not exact at any finite temperature, is restored in the infrared with power-law corrections.
The leading corrections in (\ref{lowT}) are of relative order $(T/\Lambda)^3$.
This is the scaling expected from a first-order perturbation of the ABJM fixed point by an irrelevant operator of dimension six, whose coupling scales as $\Lambda^{-3}$.
These corrections cannot, however, be attributed to the single-trace operator $\mathcal{O}_\phi$ dual to $\phi_+$ alone.
At the fixed point, $\phi_+=0$ solves the field equations at any temperature, so the thermal one-point function of $\mathcal{O}_\phi$ vanishes and its first-order contribution to the free energy is zero; a perturbation by $\mathcal{O}_\phi$ alone would therefore produce corrections of relative order $(T/\Lambda)^6$.
The $(T/\Lambda)^3$ terms instead indicate that the irrelevant deformation generated along the flow contains a dimension-six operator with non-vanishing thermal expectation value at the fixed point.
A natural candidate is a double-trace operator built from the stress tensor, such as $T_{ij}T^{ij}$, whose thermal expectation value factorizes at large $N$ and scales as $T^6$; a precise identification is left for future work.
The mechanism can nevertheless be made explicit.
At first order in $r_s^{-3}$, i.e.\ in $x^3$, the metric (\ref{hairy}) is AdS-Schwarzschild with the same $m$, written in coordinates related by the radial diffeomorphism $r\to r+(r^4/4-ml^2r/2)/r_s^3$ and by rescalings of $t$ and of $(x,y)$ by the factors $1-ml^2/(2r_s^3)$ and $1+ml^2/(4r_s^3)$.
These rescalings account for the entire first-order change of the equation of state, $s/T^2\propto1+\frac32\,ml^2/r_s^3$, with $ml^2/r_s^3=\sinh\alpha\simeq x^3/2$.
The leading correction is therefore a state-dependent rescaling of the boundary coordinates, which is how a deformation by $T_{ij}T^{ij}$ acts \cite{Taylor:2018xcy,Hartman:2018tkw}, rather than an effect of $\langle\mathcal O_\phi\rangle$.

The scalar one-point function (\ref{holoresults}) has two instructive limits.
On the lower branch, $\langle\mathcal{O}_\phi\rangle\to0$ as $T\to0$: the zero-temperature domain wall has a vanishing condensate in our scheme, even though $\varphi_{(3)}$ remains finite, and the condensate is a purely thermal effect.
On the upper branch $\langle\mathcal{O}_\phi\rangle$ grows as $e^{\alpha}$ as $T\to0$, together with the entropy density.
The undeformed ABJM plasma, i.e.\ pure AdS-Schwarzschild ($\phi_+=0$ everywhere and $C_t=C_t^{(0)}$), corresponds to $\alpha\to0$ at fixed $m$, i.e.\ to $r_s\to\infty$, $\varphi_{(0)}\to-\infty$ and $\Lambda/T\to\infty$: the scale of the deformation is pushed to infinity relative to the temperature, and the irrelevant couplings of the IR description vanish.

%%--------------------------------------------------------------------
\subsection{Comparison with non-conformal branes}
\label{sec:KScomparison}
%%--------------------------------------------------------------------

The UV regime $\alpha\to\infty$ is controlled by the single exponential $e^{-3\phi_+/\sqrt7}$ of the potential, i.e.\ by a theory with the generalized conformal structure of non-conformal branes \cite{Kanitscheider:2008kd,Kanitscheider:2009mn}.
For flat boundary sources and $D_{bdy}$ boundary spacetime dimensions, the structure is labelled by one parameter $\sigma$ and implies the dilatation Ward identity $\langle T^i{}_i\rangle=-(D_{bdy}-2\sigma)\langle\mathcal{O}\rangle$, where $\langle\mathcal{O}\rangle=-p$ in the normalization of \cite{Kanitscheider:2009mn}, together with
\be
\label{KSrelations}
\frac{\varepsilon}{p}=2\sigma-1\,,\qquad
s\propto T^{2\sigma-1}\,,\qquad
c_s^2=\frac{1}{2\sigma-1}\,,\qquad
\frac{\zeta}{\eta}=\frac{2(2\sigma-D_{bdy})}{(D_{bdy}-1)(2\sigma-1)}=2\Big(\frac{1}{D_{bdy}-1}-c_s^2\Big)\,.
\ee
With $D_{bdy}=3$, the exponent $s\propto T^{-7}$ of the upper branch fixes $2\sigma-1=-7$, i.e.\ $\sigma=-3$, and the remaining relations are then predictions.
We have checked all of them against the exact solution for $\alpha\to\infty$:
\begin{center}
\begin{tabular}{lcc}
\hline
 & this model, $\alpha\to\infty$ & (\ref{KSrelations}) with $\sigma=-3$ \\
\hline
$\varepsilon/p$ from (\ref{holoresults}) & $-7$ & $-7$ \\
$\rmd\ln s/\rmd\ln T$ & $-7$ & $-7$ \\
$c_s^2$ from (\ref{cs2}) & $-1/7$ & $-1/7$ \\
$\zeta/\eta$ from (\ref{zeta}) & $9/7$ & $9/7$ \\
$\langle T^i{}_i\rangle/p$ from (\ref{trace}) & $9$ & $D_{bdy}-2\sigma=9$ \\
\hline
\end{tabular}
\end{center}
The value $\sigma=-3$ also reproduces the hyperscaling-violating exponent, since for $z=1$ the scaling $s\propto T^{D_{bdy}-1-\theta}$ gives $\theta=D_{bdy}-2\sigma=9$.
At the other end, the IR fixed point has $c_s^2=1/2$, i.e.\ $\sigma=D_{bdy}/2$, which is the conformal case.
The flow therefore interpolates between two generalized conformal structures, and for any $\alpha\neq\alpha_*$ one can define $\sigma_{\rm eff}$ by $2\sigma_{\rm eff}-1\equiv1/c_s^2=(4\coth\alpha+3)/(2\coth\alpha-3)$, which increases from $D_{bdy}/2$ at $\alpha\to0$, diverges at $\alpha_*$, where $c_s^2=0$, and on the upper branch increases from $-\infty$ to $-3$ as $\alpha\to\infty$.

Two caveats should be stated.
First, the agreement is a test of the scaling weights, in particular of $\beta_\infty$ in (\ref{Ward}), and not of finite counterterms or of the normalization of $\langle\mathcal{O}_\phi\rangle$, because the relations (\ref{KSrelations}) follow from scale covariance alone.
Second, the derivation of \cite{Kanitscheider:2009mn} by dimensional reduction of $(2\sigma+1)$-dimensional gravity assumes $\sigma>D_{bdy}/2$ and the authors warn against its use for $\sigma<D_{bdy}/2$, and the explicit counterterms of \cite{Kanitscheider:2008kd} are given for $1<\sigma<2$.
The value $\sigma=-3$ lies outside both ranges.
The agreement above is therefore not an application of those results but a check that the UV regime of this model obeys the same relations, and it is consistent with, but does not prove, the dictionary assumptions of Section~\ref{sec:dictionary}.
The same reasoning explains the Buchel-bound results of Section~\ref{sec:hydro}: the bound is saturated for $\sigma=\mathrm{const}$ \cite{Kanitscheider:2009mn}, hence as $\alpha\to\infty$, while at finite $\alpha$ the ratio $7/(4\coth\alpha+3)<1$ measures the departure from a single generalized conformal structure.

%%%%%%%%%%%%%%%%%%%%%%%%%%
\section{A scale-invariant boundary condition and the conformal fluid}
\label{sec:multitrace}
%%%%%%%%%%%%%%%%%%%%%%%%%%

The Dirichlet theory is not the only boundary theory that can be associated with the solutions (\ref{hairy}).
In this section we show that a second, scale-invariant boundary condition, which fixes $\alpha$ instead of the source, selects the same solutions as thermal states and gives them an exactly conformal equilibrium thermodynamics.
We use the conventions of Section~\ref{sec:holofirstlaw}, in which $\rmd\hat\varepsilon=\hat T\,\rmd\hat s-\hat{\mathcal O}\,\rmd\varphi_{(0)}$.

\subsection{The boundary condition}

The scaling isometry (\ref{scaling}) maps solutions with the same $\alpha$ into one another.
On the boundary data it acts as a constant Weyl rescaling of $\hat g_{(0)}$ combined with the shift $\varphi_{(0)}\to\varphi_{(0)}-\varphi_{(1)}\ln\lambda$ (Section~\ref{sec:ward}), under which $\hat{\mathcal O}$ and $\Lambda^3$ both scale with weight three.
A Dirichlet condition breaks this symmetry, since it fixes $\Lambda$.
A boundary condition that relates the condensate to the source and preserves the symmetry must relate two quantities of weight three.
Since $\Lambda^3=l^{-3}e^{-\sqrt7\varphi_{(0)}/2}$ is the only function of $\varphi_{(0)}$ with this weight, the condition is
\be
\label{mtBC}
\hat{\mathcal O}=\frac{\rmd\hat{\mathcal W}}{\rmd\varphi_{(0)}}\,,\qquad
\hat{\mathcal W}(\varphi_{(0)})=\hat k\,\Lambda^3=\frac{\hat k}{l^3}\,e^{-\sqrt7\varphi_{(0)}/2}\,,
\ee
with a constant $\hat k$.
On the solutions, (\ref{holoresults}) converted by (\ref{hatconv}) gives $\hat{\mathcal O}=\frac{3\sqrt7}{2}(\cosh\alpha-1)\,l^4\Lambda^3/(16\pi G_4)$, so that (\ref{mtBC}) holds for every solution with a given $\alpha$, with
\be
\label{kalpha}
\hat k(\alpha)=-\frac{3l^4}{16\pi G_4}\,(\cosh\alpha-1)\,.
\ee
Since $\hat k$ decreases monotonically with $\alpha$, the condition with a fixed $\hat k\le0$ selects exactly the solutions with one value of $\alpha$, while for $\hat k>0$ there are no black-brane states.
In the language of AdS/CFT, (\ref{mtBC}) is a multi-trace boundary condition \cite{Witten:2001ua,Henneaux:2006hk,Papadimitriou:2007sj}: the generating functional is the Legendre transform of the Dirichlet one with respect to $\varphi_{(0)}$, supplemented by $\int\rmd^3\hat x\,\sqrt{-\hat g_{(0)}}\,\hat{\mathcal W}(\varphi_{(0)})$.
Because $\hat{\mathcal W}$ is cubic in the scale $\Lambda$, the construction is analogous to the conformally invariant triple-trace deformations of \cite{Hertog:2004dr,Hertog:2004rz,Hertog:2004ns}, with the important difference discussed in Section~\ref{sec:mtstatus}.

The value (\ref{kalpha}) refers to the scheme of Section~\ref{sec:onept}.
The finite counterterm (\ref{scheme}) is proportional to $\sqrt{-\gamma}\,e^{-\sqrt7\phi_+/2}$, which in the boundary coordinates (\ref{bdycoords}) evaluates to a multiple of $\Lambda^3$, the same function as $\hat{\mathcal W}$.
A change of scheme therefore shifts $\hat k$ by a constant and leaves all statements below unchanged.

\subsection{Thermodynamics}

With (\ref{mtBC}) the energy density and pressure acquire the boundary term,
\bea
\label{epsk}
&&\hat\varepsilon_k=\hat\varepsilon+\hat{\mathcal W}=\frac{l^4\Lambda^3}{16\pi G_4}\,4\sinh\alpha=\Big(\frac{r_s}{l}\Big)^{9/4}M\,,\nonumber\\
&&\hat p_k=\hat p-\hat{\mathcal W}=\tfrac12\,\hat\varepsilon_k\,,\qquad
\hat f_k=-\hat p_k=\hat f+\hat{\mathcal W}\,.
\eea
In the coordinates of Section~\ref{sec:thermo} this reads $\varepsilon_k=M$: the bulk mass density is the energy density of the fixed-$\hat k$ theory, and the bulk free energy $M-TS$ is its free energy.
The first law takes the form
\be
\label{firstlawk}
\rmd\hat\varepsilon_k=\hat T\,\rmd\hat s+\Lambda^3\,\rmd\hat k\,,
\ee
which reduces to $\rmd\hat\varepsilon_k=\hat T\,\rmd\hat s$ at fixed $\hat k$, and the stress tensor is traceless, $-\hat\varepsilon_k+2\hat p_k=-\beta_\infty\hat{\mathcal O}-3\hat{\mathcal W}=0$ by (\ref{Ward}) and (\ref{mtBC}).

The thermal states of a fixed theory have fixed $\alpha$, so that $\hat T/\Lambda$ is fixed and the scale $\Lambda$ tracks the temperature; at fixed $\alpha$, $\hat T\propto m^{7/12}$, so that every temperature is realized exactly once and there is no maximal temperature.
The equation of state is that of a conformal fluid,
\be
\label{eosk}
\hat s=\frac{\pi^3l^2}{72}\,C_t\,\hat T^2\,,\qquad
\hat\varepsilon_k=\frac23\hat T\hat s=2\hat p_k\,,
\ee
with $C_t$ fixed by $\hat k$ through (\ref{kalpha}), and the specific heat is positive.
The generalized first law (\ref{genfirstlaw}) acquires a boundary interpretation: converted to the boundary coordinates,
\be
\label{Ctk}
\Big(\frac{r_s}{l}\Big)^{9/4}\mu\,\rmd C_t=\Lambda^3\,\rmd\hat k\,,
\ee
so that the bulk work term of the thermodynamic central charge is the work term of the multi-trace coupling, and $C_t=\frac{32l^2}{\pi G_4}e^{3\alpha/2}$ is a function of $\hat k$.
In the Dirichlet theory, by contrast, $C_t$ varies along the thermal path, eq.~(\ref{Ctderived}).
For the same solution the free energies of the two theories differ by $\hat f-\hat f_k=-\hat{\mathcal W}=3(\cosh\alpha-1)\,l^4\Lambda^3/(16\pi G_4)\ge0$.

\subsection{Hydrodynamics}

Along the thermal states of fixed $\hat k$ the horizon value $\phi_+(r_h)=\frac{\sqrt7}{2}\alpha$ is constant.
The speed of sound is therefore $c_s^2=\rmd\ln\hat T/\rmd\ln\hat s=1/2$, the Eling--Oz formula \cite{Eling:2011ms}, now with the derivative taken at fixed $\hat k$, gives $\zeta=0$, and $\eta/s=1/(4\pi)$ as before.
These are the hydrodynamic data of a conformal fluid at every temperature and every $\hat k$.
Table~\ref{tab:twotheories} compares the two theories.

\begin{table}[t]
\centering
\renewcommand{\arraystretch}{1.25}
\begin{tabular}{lcc}
\hline
 & Dirichlet theory & fixed-$\hat k$ theory \\
\hline
fixed boundary data & $\varphi_{(0)}$, i.e.\ $\Lambda$ & $\hat k$, i.e.\ $\alpha$ \\
thermal path & $\alpha$ varies at fixed $\Lambda$ & $\Lambda\propto\hat T$ at fixed $\alpha$ \\
energy density & $\hat\varepsilon$, eq.~(\ref{epsM}) & $\hat\varepsilon_k=(r_s/l)^{9/4}M$ \\
temperatures & $\hat T\le T_{\max}\simeq0.19\,\Lambda$ & all $\hat T>0$ \\
$c_s^2$ & eq.~(\ref{cs2}) & $1/2$ \\
$\zeta/\eta$ & eq.~(\ref{zeta}) & $0$ \\
$\langle\hat T^i{}_i\rangle$ & $-\beta_\infty\hat{\mathcal O}$ & $0$ \\
energy bound & --- & unbounded below \\
\hline
\end{tabular}
\caption{The two boundary theories associated with the solutions (\ref{hairy}).}
\label{tab:twotheories}
\end{table}

\subsection{Status of the fixed-\texorpdfstring{$\hat k$}{k} theory}
\label{sec:mtstatus}

Two properties limit the interpretation of the fixed-$\hat k$ theory.
First, following Hertog and Horowitz \cite{Hertog:2004ns}, the energy of static configurations obeying (\ref{mtBC}) is bounded below if and only if $\mathcal V(\Lambda)=-\int\hat{\mathcal O}_{\rm sol}\,\rmd\varphi_{(0)}+\hat{\mathcal W}$ is, where $\hat{\mathcal O}_{\rm sol}$ is the condensate of the zero-temperature domain walls.
That condensate vanishes in our scheme (Section~\ref{sec:boundary}), so that $\mathcal V=\hat k\Lambda^3$, which for $\hat k<0$ is unbounded below as $\Lambda\to\infty$.
Under assumption (A3), the fixed-$\hat k$ theory therefore has no ground state, and its thermal states are at best metastable, with decay requiring large excursions of $\Lambda$.
Second, and more fundamentally, (\ref{mtBC}) treats $\varphi_{(0)}$ as a dynamical variable, which in AdS requires the source mode to be normalizable \cite{Breitenlohner:1982jf,Klebanov:1999tb,Witten:2001ua}.
Here it is not (Section~\ref{sec:dictionary}).
The condition (\ref{mtBC}) therefore defines a consistent equilibrium ensemble of the solutions, but whether it defines a dynamical boundary theory, in which $\varphi_{(0)}$ fluctuates, is not established; in particular the hydrodynamic data above are equilibrium statements.
The Dirichlet theory requires neither caveat.
We therefore regard it as the boundary theory dual to the solutions, and the fixed-$\hat k$ theory as an alternative whose equilibrium thermodynamics accounts for the conformal relations $3M=2TS$ and $\varepsilon_k=2p_k$ satisfied at fixed $\alpha$.

%%%%%%%%%%%%%%%%%%%%%%%%%%%%%%%%%%%%%%%%%%%%%%%%%%%%%%%%%%%%%%%%%%
\section{Discussion}
\label{sec:discussion}
%%%%%%%%%%%%%%%%%%%%%%%%%%

We have studied an exact static planar hairy black hole of Einstein gravity minimally coupled to a neutral scalar, which arises in a consistent truncation of type IIA supergravity on AdS$_4\times\mathbb{CP}^3$ \cite{Gauntlett:2009zw,Bak:2010yd}.
The solution itself was obtained in \cite{Cadoni:2011sm,Cadoni:2012uf}; its ABJM origin provides the dual interpretation developed here, as a minimal holographic model of thermal ABJM dynamics in which the leading thermal mode surviving the neutral truncation condenses.

The main results are the following.
On the gravity side, the two-parameter space of solutions carries a thermodynamic central charge $C_t$, a generalized first law $\rmd M = T\,\rmd S + \mu\,\rmd C_t$, and Euler and Smarr relations, organized by the scaling isometry (\ref{scaling}) and parallel to the holographic thermodynamics of CFTs with variable central charge \cite{Cong:2021fnf,Ahmed:2023snm,Gong:2023hgj,Zeyuan:2021uol}.
Holographic renormalization with the domain-wall superpotential as counterterm gives the one-point functions (\ref{holoresults}) in closed form, the relation (\ref{epsM}) between the energy density and the bulk mass density, and the Ward identity (\ref{Ward}); because the asymptotic region is not AdS, these results rest on the working assumptions of Section~\ref{sec:dictionary}.
In the dual theory with fixed source, $\alpha$ measures the temperature in units of the scale $\Lambda$, the holographic first law (\ref{holofirstlaw}) contains the work term of the source, the Ward identity is the boundary image of the bulk Smarr relation, and $C_t$ becomes a temperature-dependent effective count of degrees of freedom.

Black branes exist in two branches below $T_{\max}\simeq0.19\,\Lambda$.
On the lower branch the plasma approaches the conformal ABJM fluid, with $c_s^2\to1/2$, $\zeta/\eta\to0$ and corrections of relative order $(T/\Lambda)^3$, which arise from a state-dependent rescaling of the boundary coordinates rather than from the condensate; this accounts for the emergent IR conformality observed in \cite{Cadoni:2012uf}, while our assessment of the stability of this branch differs from the one given there for the reasons explained in Section~\ref{sec:fixedtheory}.
On the upper branch the thermodynamics is controlled by the hyperscaling-violating UV, with $c_s^2\to-1/7$ and $\zeta/\eta\to9/7$, consistent with the relations of non-conformal branes \cite{Kanitscheider:2009mn}, and at finite $\alpha$ the bulk viscosity lies below Buchel's bound.
Finally, the same solutions are the thermal states of a second boundary theory, defined by the scale-invariant multi-trace boundary condition (\ref{mtBC}), in which the energy density is the bulk mass density, the equilibrium thermodynamics is exactly conformal, and the bulk generalized first law becomes the first law with a varying multi-trace coupling, eq.~(\ref{Ctk}).
Since its energy is unbounded below and its source mode is not normalizable, this theory is established only at the level of equilibrium thermodynamics, and the Dirichlet theory is the natural dual of the solutions.
Taken together, the solution provides an exactly solvable example of a holographic RG flow triggered by a classically marginal operator, cf.\ \cite{Park:2021}, from a non-conformal UV to a conformal IR.

It is natural to ask whether the solution is also dynamically stable against linearized perturbations.
Thermodynamic stability, i.e.\ positive specific heat (\ref{heatcap}), holds on the lower branch; by the correlated stability conjecture \cite{Gubser:2000ec,Gubser:2000mm} this is a necessary, though not sufficient, condition for dynamical stability.
A systematic search for growing homogeneous modes in the coupled scalar-gravity channel (Appendix~\ref{app:dynamics}) finds none in the region of the complex frequency plane examined.
On the upper branch, by contrast, $c_s^2<0$ implies that the hydrodynamic sound mode $\omega\simeq\pm c_s k$ is unstable at sufficiently small momentum, so dynamical stability can at most hold on the lower branch.
A systematic analysis across the full parameter space is left for future work.

In the limit $\alpha\to\infty$ of a fixed boundary theory, $T\to0$ on the upper branch while $s\propto T^{-7}$ diverges; the diverging scalar field signals the need for stringy corrections in this regime.
The limit $m\to0$, $\alpha\to\infty$ with $m^{2/3}e^{\alpha/2}$ fixed keeps $S$ finite as $T\to0$, but it also sends $r_s\to0$, i.e.\ $\Lambda\to0$, and thus changes the boundary theory; at fixed $\Lambda$ it corresponds to $s/T^2=\frac{\pi^3l^2}{72}C_t\to\infty$ on the upper branch rather than to a ground-state degeneracy.

Several questions remain open.
It would be interesting to compute the holographic entanglement entropy of strip regions through the Ryu--Takayanagi formula \cite{Ryu:2006bv}.
It would provide an independent probe of the crossover between the hyperscaling-violating UV, where the entanglement entropy departs from the area law in a $\theta$-dependent way \cite{Dong:2012se}, and the ABJM fixed point in the IR, complementing the c-function (\ref{cbeta}).
The precise identification of the dual operator $\mathcal{O}_\phi$ within the full ABJM Kaluza--Klein spectrum \cite{Gauntlett:2009zw,Bak:2010yd}, and of the irrelevant deformations controlling the approach to the IR fixed point, remain interesting open problems, as does a holographic dictionary adapted to the asymptotics (\ref{UVmetric}), which would put the working assumptions of Section~\ref{sec:dictionary} on a firmer footing and decide whether the fixed-$\hat k$ boundary condition defines a dynamical theory.

We have considered only neutral, static black branes.
A further direction is the extension to charged and rotating configurations \cite{Gong:2023hgj}.
A natural extension adds a $U(1)$ gauge field with the coupling $e^{3\phi_+/\sqrt7}F^2$, which is the coupling of the Ramond-Ramond one-form to $\phi_+$ in the four-dimensional action of \cite{Bak:2010yd,Yun:2010} when the second scalar $\phi_-$ vanishes,
\be
\mathcal{L}=\frac{\sqrt{-g}}{16\pi G_4}\bigg[\mathcal{R} - \frac{1}{2}(\nabla\phi_+)^2 - V(\phi_+) - e^{3\phi_+/\sqrt{7}} F^{\mu\nu}F_{\mu\nu} \bigg]\,,
\ee
whose solution would have a richer phase structure connected to holographic superconductors \cite{Bak:2010ry}.
A truncation retaining this vector alone is not consistent, because in that action the vector has a St\"uckelberg mass term and its field strength sources $\phi_-$, so that (\ref{action}) supplemented by this Maxwell term should be regarded as a model motivated by the truncation.
Furthermore, one may investigate gravity theories with more scalar fields.
An example is the truncation to two neutral scalars $\phi_+$ and $\phi_-$,
\be
\mathcal{L}=\frac{\sqrt{-g}}{16\pi G_4}\bigg[\mathcal{R}-\frac{1}{2}(\nabla\phi_+)^2-\frac{1}{2}(\nabla\phi_-)^2 - V(\phi_+,\phi_-)\bigg]\,,
\ee
where
\be
V(\phi_+,\phi_-) = -\frac{12}{l^2}e^{-3\phi_+/\sqrt{7}+\phi_-/\sqrt{21}} + \frac{3}{2l^2}e^{-3\phi_+/\sqrt{7}+8\phi_-/\sqrt{21}} + \frac{9}{2l^2}e^{-\sqrt{7}\phi_+}.
\ee
The energy density functional of $\phi_-$ fluctuations in the $\phi_+$-hairy black brane background is \cite{Yun:2010}, to quadratic order and dropping the non-negative gradient terms along $t$, $x$ and $y$,
\bea
&&\sqrt{-g}\bigg[-g^{tt}(\partial_t\phi_-)^2+g^{rr}(\partial_r\phi_-)^2 + g^{xx}(\partial_x\phi_-)^2+g^{yy}(\partial_y\phi_-)^2 \nonumber \\
&&\quad+2V(\phi_+,\phi_-)-2V(\phi_+)\bigg] \approx\sqrt{-g}\bigg[g^{rr}(\partial_r\phi_-)^2 + \frac{4}{l^2}e^{-3\phi_+/\sqrt{7}}\phi_-^2\bigg]\geq 0\,,\nonumber
\eea
confirming stability against $\phi_-$ fluctuations.
The same conclusion holds for small simultaneous fluctuations of $\phi_-$ and $\chi$ in the further truncation,
\bea
\mathcal{L}&=&\frac{\sqrt{-g}}{16\pi G_4}\bigg[ \mathcal{R}-\frac{1}{2}(\nabla\phi_+)^2-\frac{1}{2}(\nabla\phi_-)^2 \nonumber\\
&&\qquad\quad- \frac{3}{2}e^{-2\phi_+/\sqrt{7}-4\phi_-/\sqrt{21}}(\nabla\chi)^2
  - V(\phi_+,\phi_-,\chi)\bigg]\,,\nonumber
\eea
where
\bea
V(\phi_+,\phi_-,\chi) &=& -\frac{12}{l^2}e^{-3\phi_+/\sqrt{7}+\phi_-/\sqrt{21}} +  \frac{3}{2l^2}e^{-3\phi_+/\sqrt{7}+8\phi_-/\sqrt{21}} \nonumber \\
&+& \frac{9}{2l^2}e^{-\sqrt{7}\phi_+}(1+\chi^2)^2 + \frac{6}{l^2}\chi^2e^{-5\phi_+/\sqrt{7}+4\phi_-/\sqrt{21}}.
\eea
Charged rotating black holes coupled to several scalar fields are of further interest and are left for future work.

%%%%%%%%%%%%%%%%%%%%%%%%%%
\section*{Acknowledgments}
%%%%%%%%%%%%%%%%%%%%%%%%%%

I am grateful to Kyung Kiu Kim for helpful discussions.

\appendix

%%%%%%%%%%%%%%%%%%%%%%%%%%
\section{Near-boundary analysis}
\label{app:holoren}
%%%%%%%%%%%%%%%%%%%%%%%%%%

In this appendix we collect the large-$r$ behavior of the quantities entering (\ref{onept_def}).
For the metric (\ref{hairy}) the induced metric on a surface of constant $r$ has $\sqrt{-\gamma}=r^2\sqrt{f}\,F^{3/4}$, with $f=r^2/l^2-2m/r$ and $F=(r/r_s)^3+e^{-\alpha}$, the unit normal has $n^r=\sqrt{f}\,F^{-3/4}$, and $K=\frac12 n^r\partial_r\ln\det\gamma$.
Up to terms that vanish as $r\to\infty$, the contributions of the Gibbons--Hawking part of the Brown--York tensor are
\bea
-\frac{\sqrt{-\gamma}}{8\pi G_4}\big(-K^t{}_t+K\big)&=&k\Big[-7\Big(\frac{r}{r_s}\Big)^3+14\sinh\alpha+3e^{-\alpha}\Big]\,,\nonumber\\
\frac{\sqrt{-\gamma}}{8\pi G_4}\big(-K^x{}_x+K\big)&=&k\Big[7\Big(\frac{r}{r_s}\Big)^3-8\sinh\alpha-3e^{-\alpha}\Big]\,,
\eea
while the counterterm contributes
\be
\frac{\sqrt{-\gamma}}{16\pi G_4}\,W_-=k\Big[7\Big(\frac{r}{r_s}\Big)^3-7\sinh\alpha-3\Big]
\ee
to $\varepsilon$ and the same quantity with opposite sign to $p$.
The divergences $\propto r^3$ cancel and no other divergent terms arise, giving (\ref{holoresults}).
Similarly, for the scalar one-point function,
\bea
-\frac{\sqrt{-\gamma}}{16\pi G_4}\,n^r\partial_r\phi_+&=&k\Big[-\frac{3\sqrt7}{2}\Big(\frac{r}{r_s}\Big)^3+3\sqrt7\sinh\alpha+\frac{3\sqrt7}{2}e^{-\alpha}\Big]\,,\nonumber\\
-\frac{\sqrt{-\gamma}}{16\pi G_4}\,W_-'(\phi_+)&=&k\Big[\frac{3\sqrt7}{2}\Big(\frac{r}{r_s}\Big)^3-\frac{3\sqrt7}{2}\sinh\alpha-\frac{3\sqrt7}{2}\Big]\,.
\eea
The finite counterterm discussed in Section~\ref{sec:onept} is controlled by
\be
\sqrt{-\gamma}\,e^{-\sqrt7\phi_+/2}=r^2\sqrt f\,F^{-1}\;\to\;\frac{r_s^3}{l}\,,
\ee
which yields the shifts (\ref{scheme}).

For the free energy we use the Euclidean action
\bea
I_E&=&-\frac{1}{16\pi G_4}\int\rmd^4x\sqrt{g}\,\Big[\mathcal{R}-\frac12(\nabla\phi_+)^2-V\Big]\nonumber\\
&&-\frac{1}{8\pi G_4}\int\rmd^3x\sqrt{\gamma}\,K+\frac{1}{8\pi G_4}\int\rmd^3x\sqrt{\gamma}\,\frac{W_-}{2}\,,
\eea
with Euclidean time of period $1/T$.
On shell the bulk Lagrangian reduces to $V$, by the trace of the Einstein equations, and $\sqrt{g}=r^2F^{3/2}$.
Evaluating $I_E$ between the horizon and a cutoff $r=R$ can be done in closed form.
The Euclidean section is smooth at $r=r_h$ when the time has period $1/T$, with $T$ the Hawking temperature of (\ref{MTS}), so there is no inner boundary term.
Writing $k=r_s^3/(16\pi G_4l^2)$, all contributions below are densities per unit coordinate area and per unit Euclidean time, so that they give $TI_E$ per unit area.
\begin{enumerate}
\item \emph{Bulk term.} Using $e^{-3\phi_+/\sqrt7}=F^{-3/2}$ and $e^{-\sqrt7\phi_+}=F^{-7/2}$ we have
\be
\sqrt g\,V=r^2F^{3/2}V=-\frac{21\,r^2}{2l^2}+\frac{9\,r^2}{2l^2F^2}\,,\qquad
\int\frac{r^2\,\rmd r}{F^2}=-\frac{r_s^3}{3F}\,,
\ee
the second relation following from $F'=3r^2/r_s^3$.
With $F(r_h)=e^{\alpha}$ and $r_h^3=2r_s^3\sinh\alpha$ the bulk term gives
\be
-\frac{1}{16\pi G_4}\int_{r_h}^{R}\!\rmd r\,\sqrt g\,V\;\to\;k\Big[\frac72\Big(\frac{R}{r_s}\Big)^3-7\sinh\alpha-\frac32e^{-\alpha}\Big]\,.
\ee
\item \emph{Gibbons--Hawking term.} With $\sqrt\gamma K=n^r\partial_r\sqrt\gamma$,
\be
-\frac{1}{8\pi G_4}\sqrt\gamma K\;\to\;k\Big[-\frac{21}{2}\Big(\frac{R}{r_s}\Big)^3+15\sinh\alpha+\frac92e^{-\alpha}\Big]\,.
\ee
\item \emph{Counterterm.} With $W_-$ of (\ref{fakeW}),
\be
\frac{1}{16\pi G_4}\sqrt\gamma\,W_-\;\to\;k\Big[7\Big(\frac{R}{r_s}\Big)^3-7\sinh\alpha-3\Big]\,.
\ee
\end{enumerate}
The terms $\propto R^3$ cancel, $\frac72-\frac{21}{2}+7=0$, and the finite parts add up to
\be
TI_E=k\big[\sinh\alpha+3e^{-\alpha}-3\big]=k\big[3\cosh\alpha-2\sinh\alpha-3\big]=-p\,,
\ee
where the last equality uses (\ref{holoresults}), $M=4k\sinh\alpha$ and $\sinh\alpha\tanh(\alpha/2)=\cosh\alpha-1$.
This is the free energy density $f=-p$ in the scheme $W=W_-$.
In the scheme (\ref{scheme}) the counterterm changes by $(c+3)k$, so $f\to f+(c+3)k$, in agreement with $p\to p-(c+3)k$.
As a cross-check, $TS=\frac32M$ from (\ref{MTS}), so the Euler relation gives $\varepsilon-TS=-\big(\frac12-\frac34\tanh\frac\alpha2\big)M=-p$, i.e.\ the Brown--York energy and the on-shell action agree.

\section{Homogeneous perturbations}
\label{app:dynamics}

Positive specific heat is necessary but not sufficient for dynamical stability (Section~\ref{sec:discussion}), and a direct analysis of the linearized dynamics is required; see \cite{Berti:2009kk} for a review of quasinormal-mode methods.
Here we report a systematic search for growing modes in one sector: homogeneous (zero-momentum) perturbations $\propto e^{-i\omega t}$ of the coupled scalar--gravity channel.
In the gauge $\delta g_{xx}=\delta g_{yy}=0$, $\delta g_{tr}=0$, the $tr$ Einstein equation gives $\delta B=h^2\phi_+'\,\delta\phi_+/(h^2)'$, and the $xx$ component of the trace-reversed Einstein equations, or equivalently the $rr$ constraint, determines $\delta A'$. The scalar equation then reduces to a single master equation,
\be
\delta\phi_+''+\Bigl(\ln\frac{a h^2}{b}\Bigr)'\delta\phi_+'
+\Bigl[\omega^2\frac{b^2}{a^2}-b^2V''-\frac{b^2h^2\phi_+'V'+h^2\phi_+'\phi_+''}{(h^2)'}+\frac{h^2(h^2)''\phi_+'^2}{\bigl((h^2)'\bigr)^2}-\phi_+'^2\Bigr]\delta\phi_+=0,
\label{eq:master}
\ee
with background quantities evaluated on (\ref{hairy}).
We have verified that~(\ref{eq:master}), together with the expressions for $\delta B$ and $\delta A'$, solves all components of the linearized Einstein equations.

The asymptotic structure of~(\ref{eq:master}) is unusual.
Since $b^2/a^2\simeq l^4/(r_s^3r)$, the frequency term dominates the static terms, which fall off as $r^{-2}$, for $r\gg r_s^3/(l^4\omega^2)$.
The solutions there behave as $\delta\phi_+\simeq r^{-7/4}e^{\pm2i\omega l^2\sqrt{r/r_s^3}}$. This reflects the infinite light-crossing time of the asymptotic region (Section~\ref{sec:solution}); with the weight (\ref{KGweight}) neither solution is normalizable.
Time-dependent perturbations can therefore carry energy into the asymptotic region, and the source/condensate split of Section~\ref{sec:asym} applies only to static data.
For the stability question this is not an obstacle.
A mode growing in time, $\mathrm{Im}\,\omega>0$, must be ingoing at the horizon, $\delta\phi_+\propto(r-r_h)^{-i\omega/(4\pi T)}$, and decay at large $r$ as $e^{2i\omega l^2\sqrt{r/r_s^3}}$.
These two conditions define an eigenvalue problem that is independent of the boundary condition imposed on $\varphi_{(0)}$, Dirichlet or (\ref{mtBC}).

We solved this problem numerically (in units $m=l=1$).
We integrated~(\ref{eq:master}) from the horizon and extracted the coefficient $\mathcal D(\omega)$ of the growing WKB branch at $r=3000$.
We then counted its zeros by the argument principle on the boundary of the region $|\mathrm{Re}\,\omega|\le3$, $0.1\le\mathrm{Im}\,\omega\le2.5$, using the reflection symmetry $\mathcal D(-\omega^*)=\mathcal D(\omega)^*$; this region extends to $\mathrm{Im}\,\omega\approx9$--$18\,T$ for the values of $\alpha$ considered.
The winding number vanishes for $\alpha=0.2$, $0.6$, $1.0$ and $1.5$, with phase increments below $0.2$ between successive sample points.
In addition, a scan along the imaginary axis, $\omega=i\Gamma$ with $0.003\le\Gamma\le1.6$ and the outer radius taken large enough for the WKB decomposition to be valid (up to $40\,r_s^3/\Gamma^2$ for small $\Gamma$), shows no sign change of $\mathcal D$ for $\alpha\in\{0.2,0.6,1.0,1.5,2.0\}$.
The numerically integrated solutions reproduce the WKB behavior quoted above at large $r$.
We therefore find no growing homogeneous modes in the region of the complex plane and for the values of $\alpha$ examined.
On the upper branch of the Dirichlet theory, the instability implied by $c_s^2<0$ appears in the sound mode at nonzero momentum and is not visible in this homogeneous sector.
This result should be regarded as a consistency check rather than a proof of dynamical stability: it is restricted to zero momentum, to a finite region of the frequency plane and to a finite set of values of $\alpha$.
Within these limits it is consistent with the dynamical stability of the lower branch of the Dirichlet theory, and with the metastability of the thermal states of the fixed-$\hat k$ theory (Section~\ref{sec:mtstatus}).
A systematic analysis across all channels, momenta and values of $\alpha$ remains open.

\end{document}